\documentclass[11pt]{article}
\usepackage{jheppub}
\usepackage{subcaption}
\usepackage{tikz}

\makeatletter

\@addtoreset{equation}{section}
\makeatother

\DeclareMathOperator{\Tr}{Tr}

\newcommand{\ri}{{\rm i}}

\def\th{\theta}

\def\cob{\delta}

\newcommand{\hf}{\frac{1}{2}}
\newcommand{\qu}{\frac{1}{4}}

\def\til#1{\widetilde{#1}}

\def\del{\partial}

\def\bra{\langle}
\def\ket{\rangle}

\def\la{\lambda}

\def\h#1{\widehat{#1}}

\def\Ga{\Gamma}
\def\al{\alpha}

\def\rt#1{\sqrt{#1}}

\begin{document}

\title{Phase Transition of Anti-Symmetric Wilson Loops\\ in 
$\boldsymbol{\mathcal{N}=4}$ SYM}

\author{Kazumi Okuyama}

\affiliation{Department of Physics, 
Shinshu University, Matsumoto 390-8621, Japan}

\emailAdd{kazumi@azusa.shinshu-u.ac.jp}

\abstract{We will argue that the 1/2 BPS Wilson loops in the anti-symmetric representations in the $\mathcal{N}=4$ 
super Yang-Mills (SYM) theory exhibit a phase transition
at some critical value of the 't Hooft coupling of order $N^2$. In the matrix model computation
of Wilson loop expectation values, this phase transition corresponds to the transition between the
one-cut phase and the two-cut phase.
It turns out that the one-cut phase is smoothly connected to the
small 't Hooft coupling regime and the $1/N$ corrections of Wilson loops in this phase
can be systematically computed from the topological recursion in the Gaussian matrix model.
}

\maketitle

\renewcommand{\thefootnote}{\arabic{footnote}}
\setcounter{footnote}{0}
\setcounter{section}{0}


\section{Introduction}\label{sec:intro}
1/2 BPS Wilson loops in $\mathcal{N}=4$ super Yang-Mills theory are very 
interesting observables, whose expectation values can be computed exactly by
a Gaussian matrix model via supersymmetric localization \cite{Erickson:2000af,Drukker:2000rr,Pestun:2007rz}.
From the viewpoint of bulk type IIB string theory
on $AdS_5\times S^5$, such Wilson loops correspond to some configuration of
strings or D-branes ending on the contour of Wilson loops on the boundary of $AdS_5$ 
\cite{Maldacena:1998im, Rey:1998ik,Drukker:2005kx,Yamaguchi:2006tq,Hartnoll:2006is,Okuyama:2006jc,Gomis:2006sb,Gomis:2006im}.
In particular, Wilson loops in the $k^{\text{th}}$ anti-symmetric representation
$W_{A_k}$ correspond to D5-branes with $k$ unit of electric flux on their worldvolume.
In the usual 't Hooft limit with an additional scaling 
\begin{align}
 N,k\to\infty~~~\text{with}\quad x=\frac{k}{N}:~\text{fixed},
\label{eq:xlim}
\end{align}
the leading term in the $1/N$ expansion of $W_{A_k}$ is successfully matched
with the DBI action of D5-branes on the bulk gravity side.
A similar matching was found between Wilson loops in the symmetric representations
and D3-branes on the bulk side.
However, a mismatch at the subleading order in the $1/N$ expansion 
was reported for both symmetric and anti-symmetric representations \cite{Faraggi:2014tna,Faraggi:2011ge,Faraggi:2011bb,Buchbinder:2014nia} (see also 
\cite{Zarembo:2016bbk} for a review of the status of this problem).
Recently, the first $1/N$ correction of Wilson loops 
was computed for both symmetric \cite{Chen-Lin:2016kkk} and anti-symmetric \cite{Gordon:2017dvy}
representations, but the mismatch still remains as an issue.

In this paper, we will consider the $1/N$ corrections of Wilson loops 
$W_{A_k}$ in the anti-symmetric representations.
To study $W_{A_k}$, it is convenient to consider
the generating function of $W_{A_k}$
\begin{align}
 P(z)=\sum_{k=0}^N z^k W_{A_k}e^{\frac{\la k^2}{8N^2}}.
\label{eq:gen-P}
\end{align}
Here, the extra factor $e^{\frac{\la k^2}{8N^2}}$
comes from the $U(1)$ part of $U(N)$
gauge theory \cite{Gordon:2017dvy} and $W_{A_k}$ in \eqref{eq:gen-P} denotes the expectation
value of
the Wilson loop in $SU(N)$ gauge theory.
This generating function has a simple expression
in the Gaussian matrix model
\begin{align}
 P(z)=\bra \det(1+ze^M)\ket,
\label{eq:P-def}
\end{align}
where the expectation value is defined by
\begin{align}
 \bra f(M)\ket=\frac{1}{Z}\int dM e^{-\frac{2N}{\la}\Tr M^2}
f(M),
\end{align}
with $\la$ being the 't Hooft coupling and $Z=\int dM e^{-\frac{2N}{\la}\Tr M^2}$. 
We would like to find the $1/N$ expansion of this generating function
\begin{align}
 \log P(z)= \sum_{n=0}^\infty N^{1-n}J_n(z).
\label{eq:log-exp}
\end{align}
In order to compute this expansion, there are two approaches.
The first approach is to regard the operator $\det(1+ze^M)$ 
as a part of the potential
of matrix integral
\begin{align}
 P(z)=\frac{1}{Z}\int dM e^{-NV(M)},\quad V(M)=\frac{2}{\la}\Tr M^2
-\frac{1}{N}\Tr\log(1+ze^M),
\label{eq:V-mod}
\end{align}
and study the large $N$ behavior of eigenvalue distribution under this modified potential.
This approach was recently considered in \cite{Gordon:2017dvy}.
The second one is to treat the insertion $\det(1+ze^M)$ as an operator in the original Gaussian
matrix model.
It turns out that these two computations lead  to the same result of $1/N$ expansion
\eqref{eq:log-exp}
as long as $\la\ll N^2$. In other words,
the small $\la$ regime and the large $\la$ regime belong to the same phase
when $\la\ll N^2$, and we can safely treat the insertion of $\det(1+ze^M)$ 
as a small perturbation of the original Gaussian matrix model.

However, when $\la\sim N^2$ we should take into account the backreaction of the operator $\det(1+ze^M)$.
As suggested in \cite{Gordon:2017dvy}, in this regime we can take a scaling limit
\begin{align}
 N,\la\to \infty~~\text{with}~~\xi=\frac{\rt{\la}}{N}:~\text{fixed}. 
\label{eq:scaling}
\end{align}
In this limit the expansion \eqref{eq:log-exp} is re-organized into a new expansion
\begin{align}
 \log P(z)= \sum_{g=0}^\infty N^{2-2g}G_g(\xi,\th),
\label{eq:closed-expansion}
\end{align}
where we have also set
\begin{align}
 z=e^{-\rt{\la}\cos\th}.
\label{eq:z-cos}
\end{align}
Notice that $\log P(z)$ is $\mathcal{O}(N)$ in \eqref{eq:log-exp}
while it is $\mathcal{O}(N^2)$ in \eqref{eq:closed-expansion}.
Moreover, one can see that the expansion \eqref{eq:closed-expansion}
has the same form as the genus expansion of {\it closed string theory}.
This strongly suggests that the Wilson loops in this scaling limit \eqref{eq:scaling}
are no longer described by D5-branes, but they are replaced by a closed string background
without D5-branes.
However, the physical interpretation of this closed string background on the bulk side
is unclear at the moment.
We will comment on some possible interpretation in section \ref{sec:discussion}.

In this scaling limit \eqref{eq:scaling}, we have one free parameter $\xi$.
It turns out that when $\xi$ is small the potential \eqref{eq:V-mod}
has only one minimum near $M=0$, but as we increase the value of
$\xi$ this potential develops a new local minimum in addition to the original one
near $M=0$ (see Fig.~\ref{fig:pot}).
Then it is natural to conjecture that there is a phase transition
between the one-cut phase
and the two-cut phase, in which the eigenvalue distribution
has one support or two disjoint supports, respectively.
We compute the eigenvalue density in the one-cut phase
in this scaling limit
and find evidence that the support of eigenvalue distribution
splits into two parts at some critical value $\xi=\xi_c$.
However, we could not determine
the precise value of the critical point $\xi_c$.
Also, the explicit construction of the two-cut solution is beyond the scope of this paper.
It would be very interesting to understand the nature of this phase transition better.
 
The rest of this paper is organized as follows.
In section \ref{sec:gen}, we compute the small $\la$ expansion
of the generating function \eqref{eq:log-exp}
and show that our result of $J_1(z)$
reproduces the result in \cite{Gordon:2017dvy}.
In section \ref{sec:top}, we compute the higher order corrections $J_n(z)$
in the $1/N$ expansion \eqref{eq:log-exp}
using the topological recursion of the Gaussian matrix model.
In section \ref{sec:large-la},
we consider the behavior of the generating function and the Wilson loops
in the scaling limit \eqref{eq:scaling}
and find evidence that there is a phase transition between the one-cut phase and
the two-cut phase at some critical value $\xi=\xi_c$.
We conclude in section \ref{sec:discussion} with some discussion of future directions.
In appendix \ref{app:small-la},
we consider the small $\la$ expansion of $W_{A_k}$.

\section{Small $\la$ expansion of the generating function}\label{sec:gen}
In this section, we will consider the $1/N$ expansion
of the generating function \eqref{eq:log-exp}
\begin{align}
 J(z)=\frac{1}{N}\log P(z)=\sum_{n=0}^\infty N^{-n}J_n(z).
\label{eq:Jz-Jn}
\end{align}
We will compute this expansion using the approach of operator insertion
in the Gaussian matrix model, as discussed in section \ref{sec:intro}.
As we will see below, our result of first correction $J_1(z)$ 
agrees with the one in \cite{Gordon:2017dvy}
obtained from the shift of potential \eqref{eq:V-mod}.

Let us first consider the small $\la$ expansion of $J(z)$
using the standard Feynman diagram expansion of Gaussian matrix model.
To do this, it is convenient to rescale the matrix variable $M\to\rt{\la}M$
and rewrite the generating function $P(z)$ in \eqref{eq:P-def} as
\begin{align}
 P(z)=\bra\det(1+ze^{\rt{\la}M})\ket,
\label{eq:Pz-rescale}
\end{align}
where the expectation value is taken in the Gaussian measure
with a potential $V(M)=2\Tr M^2$
\begin{align}
 \bra f(M)\ket=\frac{1}{Z}\int dM e^{-2N\Tr M^2}f(M).
\label{eq:Gauss}
\end{align}
Note that in this normalization the eigenvalues
of $N\times N$ hermitian matrix $M$ 
are distributed along the cut $[-1,1]$
in the large $N$ limit.

Then the small $\la$ expansion of $J(z)$ is easily obtained by expanding
$\det(1+ze^{\rt{\la}M})$ around $\la=0$.
For instance, the order $\mathcal{O}(\la)$ term in the small $\la$ expansion can be found as
\begin{align}
 \begin{aligned}
  J(z)&=\frac{1}{N}\log\bra \det(1+ze^{\rt{\la}M})\ket=
\frac{1}{N}\log\bra \det(1+z+z\rt{\la}M+\hf z\la M^2+\cdots)\ket\\
&=\log(1+z)+\frac{z\bra \Tr M^2\ket+z^2 \bra (\Tr M)^2\ket}{(1+z)^2}\frac{\la}{2N}
+\mathcal{O}(\la^2).
\label{eq:J-order-la}
\end{aligned}
\end{align}
Using the propagator of Gaussian matrix model \eqref{eq:Gauss}
\begin{align}
 \bra M_{ab}M_{cd}\ket=\frac{1}{4N}\cob_{ad}\cob_{bc},
\end{align}
the correlators  appearing in \eqref{eq:J-order-la}
are given by
\begin{align}
 \bra \Tr M^2\ket=\frac{N}{4},\quad
\bra (\Tr M)^2\ket =\qu.
\end{align}
In this manner we can easily find the first few terms of small $\la$ expansion
\begin{align}
 \begin{aligned}
  J(z)&=\log(1+z)+\frac{Nz+z^2}{8(1+z)^2}\frac{\la}{N}\\
&+\Bigl[2N^2(1-4z+z^2)-6Nz(2z-3)+(1-4z+13z^2)\Bigr]\frac{z}{384(1+z)^4}\frac{\la^2}{N^2}+
\mathcal{O}(\la^3). 
\label{eq:Jla-order2}
\end{aligned}
\end{align} 
However, the number of diagrams grows rapidly at the higher order of small
$\la$ expansion, and this Feynman diagramatic approach is not so 
useful in practice.

To compute the higher order terms of small $\la$ expansion,
we can use the exact form of the generating function $P(z)$
found in \cite{Fiol:2013hna}
\begin{align}
 P(z)=\det(1+zA),
\label{eq:P-exact}
\end{align}
where $A$ is an $N\times N$ matrix whose $(i,j)$ component is given by
\begin{align}
 A_{i,j}=L^{(j-i)}_{i-1}\Bigl(-\frac{\la}{4N}\Bigr)e^{\frac{\la}{8N}},
\end{align}
and $L_n^{(a)}(x)$ denotes the generalized Laguerre polynomial.
From \eqref{eq:Jla-order2}, we observe that the coefficient of
$(\la/N)^n$  is an $n^{\text{th}}$ order polynomial of $N$.
Then the easiest way to find this polynomial is to expand the exact expression \eqref{eq:P-exact}
for small values of $N$ $(N=1,2,\cdots,n+1)$ 
and plug the result into the function `{\tt InterpolatingPolynomial}' in
{\tt Mathematica}.

Using this method we can find the small $\la$ expansion of $J(z)$
up to very high order. Then we can extract the $1/N$ corrections $J_n(z)$
in \eqref{eq:Jz-Jn}  quite easily: 
\begin{align}
 \begin{aligned}
  J_0(z)&=\log(1+z)+\frac{z}{8(1+z)^2}\la+\frac{z(1-4z+z^2)}{192(1+z)^4}\la^2+
\frac{z \left(z^4-26 z^3+66 z^2-26 z+1\right)}{9216
   (z+1)^6}\la^3\\
&\quad +\frac{z \left(z^6-120 z^5+1191 z^4-2416 z^3+1191 z^2-120
   z+1\right)}{737280 (z+1)^8}\la^4+\mathcal{O}(\la^5), \\
J_1(z)&=\frac{z^2}{8 (z+1)^2}\la-\frac{z^2 (2 z-3)}{64 (z+1)^4}\la^2
-\frac{z^2 \left(z^3-15 z^2+23 z-5\right)}{768 (z+1)^6}\la^3\\
&\quad -\frac{z^2 \left(2 z^5-147 z^4+1048 z^3-1558 z^2+558
   z-35\right)}{73728 (z+1)^8}\la^4+\mathcal{O}(\la^5),\\
J_2(z)&= \frac{z \left(1-4z+13 z^2\right)}{384 (z+1)^4}\la^2
+\frac{z \left(19 z^4-170 z^3+168 z^2-26 z+1\right)}{4608
   (z+1)^6}\la^3\\
&\quad +\frac{z \left(25 z^6-1176 z^5+6231 z^4-7216 z^3+2079
   z^2-120 z+1\right)}{147456 (z+1)^8}\la^4+\mathcal{O}(\la^5),\\
J_3(z)&=-\frac{z^2 \left(5 z^3-30 z^2+21 z-4\right)}{1536
   (z+1)^6}\la^3\\
&\quad -\frac{z^2 \left(28 z^5-903 z^4+3776 z^3-3650 z^2+948
   z-55\right)}{73728 (z+1)^8}\la^4+\mathcal{O}(\la^5).
\end{aligned}
\label{eq:smallla-J}
\end{align}
We can easily check that the small $\la$ expansion of
$J_0(z)$
in \eqref{eq:smallla-J}
is reproduced from the following expression
\begin{align}
 J_0(z)=\frac{2}{\pi}\int_{-1}^1du\rt{1-u^2}f(u,z),
\label{eq:J0-wigner}
\end{align}
where we introduced the notation
\begin{align}
 f(u,z)=\log(1+ze^{\rt{\la}u}).
\label{eq:def-fu}
\end{align}
This expression \eqref{eq:J0-wigner}
is just the planar expectation value
of the operator $f(M,z)$
in the semi-circle distribution of Gaussian matrix model.

Also, one can easily check that the small $\la$ expansion of $J_1(z)$
in \eqref{eq:smallla-J} is reproduced from the following expression
\begin{align}
 J_1(z)=\frac{1}{8\pi^2}\int_{-1}^1du \int_{-1}^1dv \frac{1-uv}{\rt{(1-u^2)(1-v^2)}}
\frac{\bigl[f(u,z)-f(v,z)\bigr]^2}{(u-v)^2}.
\label{eq:J2-fvar}
\end{align}
According to the result in \cite{Haagerup}, this expression \eqref{eq:J2-fvar} 
is nothing but the connected two-point function of
the operator $f(M,z)$ in the {\it Gaussian matrix model}  
\begin{align}
 J_1(z)=\hf \bra f(M,z)f(M,z)\ket_{\text{conn}}.
\label{eq:J1-closed}
\end{align} 

One can show that our result \eqref{eq:J2-fvar} is equivalent to the
result in \cite{Gordon:2017dvy}, as follows.
Using the relation
\begin{align}
 \frac{1-uv}{(u-v)^2\rt{1-v^2}}=\del_v\frac{\rt{1-v^2}}{u-v},
\label{eq:del-rhouv}
\end{align}
and integration by parts, we can rewrite \eqref{eq:J2-fvar} as
\begin{align}
 J_1(z)=\frac{1}{4\pi^2}\int_{-1}^1 du \rho_1(u)f(u,z),
\label{eq:J1-rho}
\end{align}
where $\rho_1(u)$ is given by
\begin{align}
 \rho_1(u)=\frac{1}{\rt{1-u^2}}\int_{-1}^1 dv \frac{\rt{1-v^2}}{u-v}\del_v f(v,z).
\end{align}
This is exactly the same as the eigenvalue density at the subleading order
obtained in \cite{Gordon:2017dvy}.
We should stress that our result is obtained from the 
perturbative computation in Gaussian matrix model
without taking into account the backreaction.
On the other hand, the computation in \cite{Gordon:2017dvy}
is based on the shift of potential \eqref{eq:V-mod} mentioned in section \ref{sec:intro}.
Interestingly, these two computations give the same result.
As we will explain in section \ref{sec:large-la}, the physical reason behind this agreement is that these two
computations are done in the same phase (one-cut phase) of matrix integral 
and hence they are smoothly connected.

In \cite{Gordon:2017dvy},
it was also noticed that the derivative of $J_1(z)$
with respect to $\la$ has a simple expression.
This can also be proved easily from our result \eqref{eq:J2-fvar} or \eqref{eq:J1-rho}.
Using the relation 
\begin{align}
 \del_\la f(u,z)=\frac{u}{2\rt{\la}}\del_u f(u,z),
\end{align}
and \eqref{eq:del-rhouv}, one can show that after integration by parts
the $\la$-derivative of \eqref{eq:J1-rho} becomes
\begin{align}
 \del_\la J_1(z)=\frac{1}{8\pi^2\la}\int_{-1}^1 du \int_{-1}^1 dv
\frac{1+uv}{\rt{(1-u^2)(1-v^2)}}\del_u f(u,z)\del_v f(v,z), 
\label{eq:dla-J1}
\end{align}
which reproduces the result in \cite{Gordon:2017dvy}.

\subsection{Numerical test}
We can numerically test the subleading
corrections of $W_{A_k}$ from the $1/N$ expansion
of $J(z)$.
Let us define the leading order (LO) and the next-to-leading order
(NLO) terms in the $1/N$ expansion of $W_{A_k}$
\begin{align}
\begin{aligned}
 W_{A_k}^{\text{LO}}&=\oint\frac{dz}{2\pi\ri z^{k+1}} e^{NJ_0(z)},\\
W_{A_k}^{\text{LO}+\text{NLO}}&=
e^{-\frac{\la k^2}{8N^2}}\oint\frac{dz}{2\pi\ri z^{k+1}} e^{NJ_0(z)+J_1(z)}.
\end{aligned}
\end{align}
Here the extra factor $e^{-\la k^2/8N^2}$ comes from the $U(1)$ part, as mentioned
in section \ref{sec:intro}.
From our result of $J_{0}(z)$ in \eqref{eq:J0-wigner}
and $\del_\la J_1(z)$ in \eqref{eq:dla-J1},
we can find the power series expansion of 
$J_{0}(z)$ and $J_{1}(z)$
\footnote{These expressions are obtained by M.~Beccaria. We would like to thank him 
for sharing his unpublished note.}
\begin{align}
 \begin{aligned}
  J_0(z)&=-\frac{2}{\rt{\la}}\sum_{m=1}^\infty \frac{(-z)^m}{m^2} I_1(m\rt{\la}),\\
J_1(z)&=\frac{1}{8}\sum_{m=1}^\infty(-z)^m \int_{0}^\la d\la' \sum_{a=1}^{m-1}
\Bigl[I_0(a\rt{\la'})I_0((m-a)\rt{\la'})+I_1(a\rt{\la'})I_1((m-a)\rt{\la'})\Bigr],
 \end{aligned}
\label{eq:J01-z}
\end{align}
where $I_\nu(x)$ denotes the modified Bessel function of the first kind.
By extracting the coefficient of $z^k$ in 
$e^{NJ_0(z)}$ and $e^{NJ_0(z)+J_1(z)}$, 
we can compute the $1/N$ correction of
$W_{A_k}$ numerically in the small $\la$ regime. 
Then we can compare them with the exact value of $W_{A_k}$
\begin{align}
 W_{A_k}^{\text{exact}}=e^{-\frac{\la k^2}{8N^2}}\oint\frac{dz}{2\pi\ri z^{k+1}}\det(1+zA).
\label{eq:W-exact}
\end{align}

\begin{figure}[htb]
\centering
\subcaptionbox{Leading term of $W_{A_k}$\label{sfig:lead}}{\includegraphics[width=6cm]{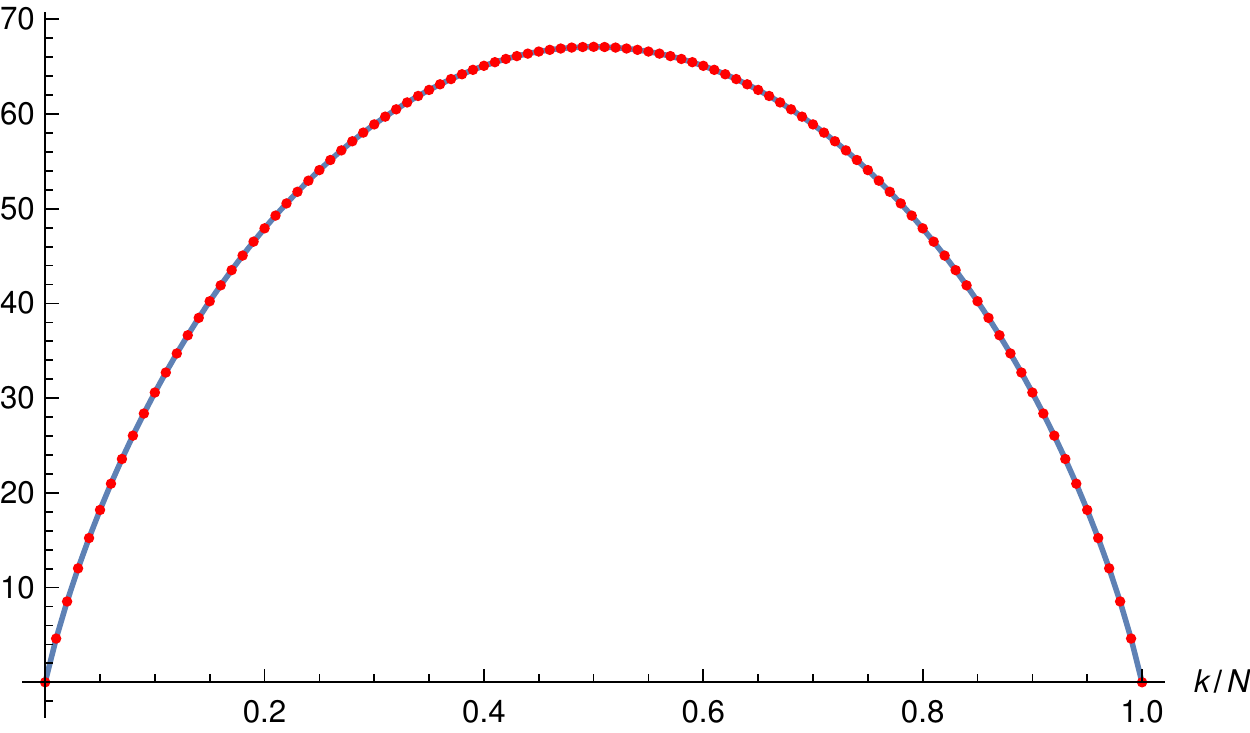}}
\hskip15mm
\subcaptionbox{Sub-leading term of $W_{A_k}$\label{sfig:sub}}{\includegraphics[width=6cm]{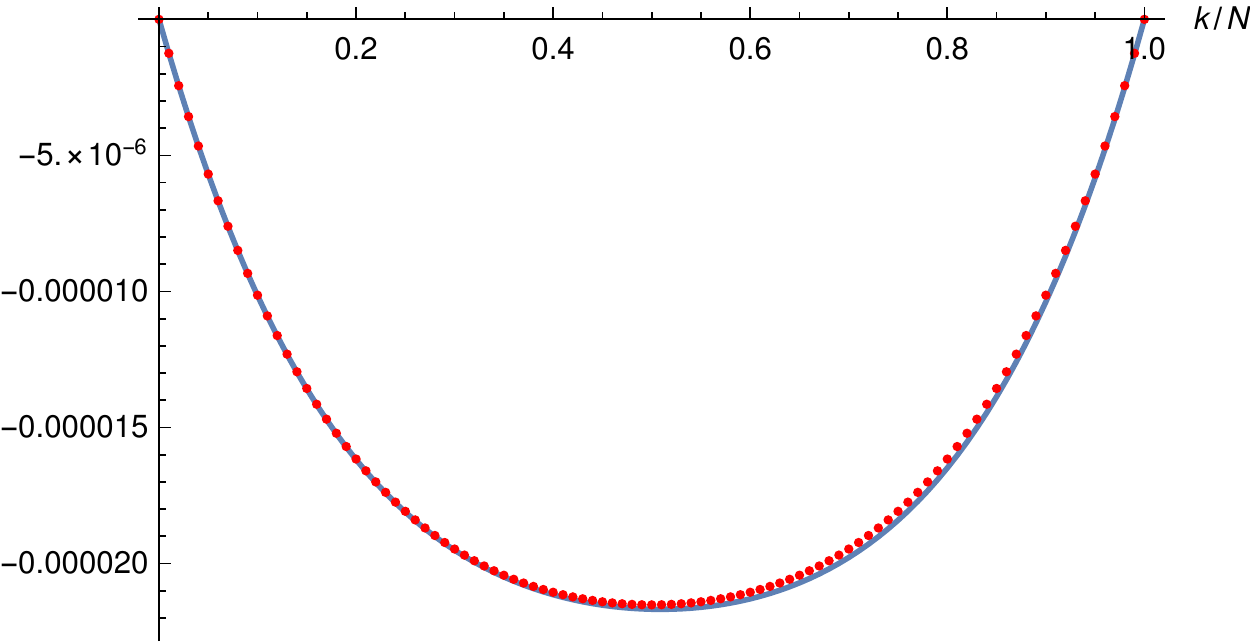}}
  \caption{
Plot 
of $W_{A_k}$ as a function of $k/N$ for $N=100,~\la=0.1$.
In \subref{sfig:lead}, the red dots represent $\log W_{A_k}^{\text{exact}}$ while
the blue curve is $\log W_{A_k}^{\text{LO}}$. In \subref{sfig:sub},
the red dots represent
$\log (W_{A_k}^{\text{exact}}/W_{A_k}^{\text{LO}})$ while the blue curve
is $\log (W_{A_k}^{\text{LO+NLO}}/W_{A_k}^{\text{LO}})$.
}
  \label{fig:Wnum}
\end{figure}

In Fig.~\ref{fig:Wnum}, we plot the leading order and 
the next-to-leading order terms in $1/N$
expansion for $N=100$ and $\la=0.1$.
As we can see from Fig.~\ref{fig:Wnum}, our result of $1/N$ correction 
\eqref{eq:J01-z} correctly reproduces
the exact result for small $\la$.

\section{$1/N$ corrections from topological recursion}\label{sec:top}
As we have seen in the previous section, the $1/N$ 
expansion of $J(z)$ in the one-cut phase can be obtained by
the perturbative computation in 
the Gaussian matrix model {\it without taking into account the backreaction}.
In general, the log of the generating function $P(z)$
in \eqref{eq:Pz-rescale} is expanded as
\begin{align}
 \begin{aligned}
\log P(z)&=  \log\bra e^{\Tr\log (1+ze^{\rt{\la}M})}\ket = \sum_{h=1}^\infty\frac{1}{h!}\left\bra \Bigl[\Tr\log (1+ze^{\rt{\la}M})
\Bigr]^h\right\ket_{\text{conn}},
 \end{aligned}
\end{align}
and the connected $h$-point function  is
expanded as
\begin{align}
 \left\bra \Bigl[\Tr\log (1+ze^{\rt{\la}M})
\Bigr]^h\right\ket_{\text{conn}}=\sum_{g=0}^\infty N^{2-2g-h}\mathcal{J}_{g,h}(z).
\end{align}
Finally, the $n^{\text{th}}$ order term in the $1/N$ 
expansion of $J(z)$ in \eqref{eq:Jz-Jn} is given by
\begin{align}
 J_n(z)=\sum_{\substack{2g+h-1=n\\g\geq0,~h\geq1}}\frac{1}{h!}\mathcal{J}_{g,h}(z). 
\label{eq:Jn-Jgh}
\end{align}
In this notation, $J_0(z)$ and $J_1(z)$ in the previous section are written as
\begin{align}
 J_0(z)=\mathcal{J}_{0,1}(z),\quad
J_1(z)=\hf \mathcal{J}_{0,2}(z).
\end{align}

To compute $\mathcal{J}_{g,h}(z)$, we observe that
they are related to the genus expansion of 
the correlator of resolvents
\begin{align}
 \Biggl\bra\prod_{i=1}^h \Tr \frac{1}{x_i-M}\Biggr\ket_{\text{conn}}=
\sum_{g=0}^\infty N^{2-2g-h}W_{g,h}(x_1,\cdots,x_h).
\end{align}
For instance, $W_{0,1}$ and $W_{0,2}$ in the Gaussian matrix model
are given by
\begin{align}
 \begin{aligned}
  W_{0,1}(x)&=\frac{2}{\pi}\int_{-1}^1 du\rt{1-u^2}\,r(u,x)=2x-2\rt{x^2-1},
\label{eq:w01}
 \end{aligned}
\end{align}
and 
\begin{align}
  \begin{aligned}
W_{0,2}(x,y)&=\frac{1}{4\pi^2}\int_{-1}^1du
\int_{-1}^1dv \frac{1-uv}{\rt{(1-u^2)(1-v^2)}}\frac{\bigl[r(u,x)-r(v,x)\bigr]
\bigl[r(u,y)-r(v,y)\bigr]}{(u-v)^2}\\
&=-\frac{1}{2(x-y)^2}\left[\frac{1-xy}{\rt{(x^2-1)(y^2-1)}}+1\right],
 \end{aligned}
\label{eq:w02}
\end{align}
where we have introduced the notation $r(u,x)$ as
\begin{align}
 r(u,x)=\frac{1}{x-u}.
\end{align}
Comparing the above expression of $W_{0,1}$ and $W_{0,2}$ with 
$\mathcal{J}_{0,1}$ in \eqref{eq:J0-wigner} and $\mathcal{J}_{0,2}$ in \eqref{eq:J2-fvar},
we can easily see the dictionary between $W_{g,h}$ and
$\mathcal{J}_{g,h}$. Namely, once we know the integral representation of $W_{g,h}$
\begin{align}
 W_{g,h}(x_1,\cdots,x_h)=
\int d^h u \,\rho_{g,h}(u_1,\cdots,u_h)\mathcal{T}_{g,h}\bigl[r(u_1,x_1),\cdots, r(u_h,x_h)\bigr],
\label{eq:wgh-density}
\end{align}
with some density $\rho_{g,h}$ and a multi-linear differential operator 
$\mathcal{T}_{g,h}$, then $\mathcal{J}_{g,h}$
is readily obtained by replacing $r(u_i,x_i)\to f(u_i,z)$
\begin{align}
 \mathcal{J}_{g,h}(z)=\int d^h u \,\rho_{g,h}(u_1,\cdots,u_h)\mathcal{T}_{g,h}\bigl[f(u_1,z),
\cdots, f(u_h,z)\bigr].
\label{eq:Jgh-density}
\end{align}

The genus-$g$, $h$-point function of resolvent in the Gaussian matrix model 
is easily obtained from the topological recursion \cite{Eynard:2004mh} (see also 
 \cite{Eynard:2008we} for a review).
For the Gaussian potential $V(M)=2\Tr M^2$
the recursion relation reads
\begin{align}
 \begin{aligned}
  4x_1W_{g,h}(x_1,\cdots,x_h)=&W_{g-1,h+1}(x_1,x_1,x_2,\cdots,x_h)+4\cob_{g,0}\cob_{h,1}\\
&+\sum_{I_1\sqcup I_2=\{2,\cdots,h\}}\sum_{g'=0}^gW_{g',1+|I_1|}(x_1,x_{I_1})
W_{g-g',1+|I_2|}(x_1,x_{I_2})
\\
&+\sum_{j=2}^h \frac{\del}{\del x_j}\frac{W_{g,h-1}(x_1,\cdots,\h{x}_j,\cdots,x_h)
-W_{g,h-1}(x_2,\cdots,x_h)}{x_1-x_j},
 \end{aligned}
\end{align}
and we can compute $W_{g,h}$ in the Gaussian matrix model recursively
starting from $W_{0,1}$ in \eqref{eq:w01}.
For $(g,h)=(0,2)$ the recursion relation is given by
\begin{align}
 4xW_{0,2}(x,y)=2W_{0,1}(x)W_{0,2}(x,y)+\frac{\del}{\del y}\frac{W_{0,1}(x)-W_{0,1}(y)}{x-y},
\end{align}
which reproduces \eqref{eq:w02}.

For $(g,h)=(1,1)$ we find
\begin{align}
 W_{1,1}(x)=\frac{W_{0,2}(x,x)}{4\rt{x^2-1}}=\frac{1}{16(x^2-1)^{5/2}}=\del_x^2\left(\frac{2x^2-1}{48\rt{x^2-1}}\right).
\label{eq:W11}
\end{align}
The density $\rho_{1,1}(u)$ is obtained by
taking the discontinuity across the real axis.
However, if we do this naively in the first expression of \eqref{eq:W11}
we would have $\rho_{1,1}(u)\sim (1-u^2)^{-5/2}$ whose integral is not convergent near $u=\pm1$.
Thus we have to perform the integration by parts and use the last expression of \eqref{eq:W11}
with $\rho_{1,1}(u)\sim (1-u^2)^{-1/2}$ whose integral is convergent
near $u=\pm1$.
In this way we can rewrite $W_{1,1}$ in \eqref{eq:W11}
into the form of the density integral in \eqref{eq:wgh-density}
\begin{align}
W_{1,1}(x)&=\int_{-1}^1du\rho_{1,1}(u)
\del_u^2 r(u,x),\qquad
\rho_{1,1}(u)=\frac{1}{48\pi}\frac{2u^2-1}{\rt{1-u^2}}.
\end{align}

In general, the density integral of $W_{g,h}$ 
in \eqref{eq:wgh-density} is obtained by using the relation
\begin{align}
\begin{aligned}
(x^2-1)^{-n-\hf}&= \frac{(-1)^n}{(2n-1)!!}\del_x^n\frac{T_n(x)}{\rt{x^2-1}},\\
x(x^2-1)^{-n-\hf}&=\frac{(-1)^n}{(2n-1)!!}\del_x^n\frac{T_{n-1}(x)}{\rt{x^2-1}},
\end{aligned}
\label{eq:del-Tn}
\end{align}
where $T_n(x)$ is the Chebychev polynomial of the first kind
satisfying
\begin{align}
 T_n(\cos\th)=\cos n\th.
\end{align}
Then we can rewrite $W_{g,h}$
as a sum derivatives of $p(x_j)/\prod_j\rt{x_j^2-1}$
with polynomial coefficient $p(x_j)$.
After the integration by parts, we can find the density 
$\rho_{g,h}$ which behaves as $(1-u^2)^{-1/2}$ near the end-point of the cut $u=\pm 1$.

For $(g,h)=(0,3)$ the recursion relation is given by
\begin{align}
 \begin{aligned}
  4xW_{0,3}(x,y,z)&=2W_{0,1}(x)W_{0,3}(x,y,z)+2W_{0,2}(x,y)W_{0,2}(x,z)\\
&+\frac{\del}{\del y}\frac{W_{0,2}(x,z)-W_{0,2}(y,z)}{x-y}+
\frac{\del}{\del z}\frac{W_{0,2}(x,y)-W_{0,2}(y,z)}{x-z},
 \end{aligned}
\end{align}
and the solution is
\begin{align}
\begin{aligned}
 W_{0,3}(x,y,z)&=\frac{1+xy+yz+zx}{8\bigl[(x^2-1)(y^2-1)(z^2-1)\bigr]^{\frac{3}{2}}}.
\end{aligned}
\label{eq:w03}
\end{align}
Using the relation \eqref{eq:del-Tn}
this is rewritten as
\begin{align}
 \begin{aligned}
  W_{0,3}(x,y,z)
&=-\frac{1}{8}\del_x\del_y\del_z\frac{x+y+z+xyz}{\rt{(x^2-1)(y^2-1)(z^2-1)}}.
\label{eq:w03}
 \end{aligned}
\end{align}

Now, from \eqref{eq:Jn-Jgh} we can compute the next order term $J_2(z)$  in the $1/N$
expansion of $J(z)$
\begin{align}
 J_2(z)=\mathcal{J}_{1,1}(z)+\frac{1}{3!}\mathcal{J}_{0,3}(z).
\end{align}
Applying
the general procedure \eqref{eq:wgh-density}
and \eqref{eq:Jgh-density}
to the result of $W_{1,1}$ in \eqref{eq:W11} and $W_{0,3}$ in \eqref{eq:w03}, 
we find
\begin{align}
 \begin{aligned}
  \mathcal{J}_{1,1}(z)&=\frac{1}{48\pi}\int_{-1}^1du\frac{2u^2-1}{\rt{1-u^2}}
\del_u^2 f(u,z),\\
\mathcal{J}_{0,3}(z)&=\frac{1}{8\pi^3}\int_{-1}^1 du\int_{-1}^1 dv\int_{-1}^1 dw 
\frac{u+v+w+uvw}{\rt{(1-u^2)(1-v^2)(1-w^2)}}\del_u f(u,z)
\del_v f(v,z)\del_w f(w,z).
 \end{aligned}
\end{align}
One can check that this reproduces the small
$\la$ expansion of $J_2(z)$ in \eqref{eq:smallla-J}.

We can push this computation to the next order term $J_3(z)$
\begin{align}
J_3(z)= \frac{1}{2}\mathcal{J}_{1,2}(z)+\frac{1}{4!}\mathcal{J}_{0,4}(z).
\end{align} 
For $(g,h)=(1,2)$, the resolvent is obtained from the topological recursion
as
\begin{align}
 \begin{aligned}
  W_{1,2}(x,y)&=\frac{5}{64}(1+xy)\left[(x^2-1)^{-3/2}(y^2-1)^{-7/2}+(x^2-1)^{-7/2}(y^2-1)^{-3/2}\right]\\
&+\frac{1}{16}\left[(x^2-1)^{-3/2}(y^2-1)^{-5/2}+(x^2-1)^{-5/2}(y^2-1)^{-3/2}\right]\\
&+\frac{3}{64}(1+xy)(x^2-1)^{-5/2}(y^2-1)^{-5/2}.
 \end{aligned}
\end{align}
Using the relation \eqref{eq:del-Tn} this is rewritten as
\begin{align}
 \begin{aligned}
 W_{1,2}(x,y)&=\frac{1}{192} \left(\del_x\del_y^3\frac{T_1(x)T_3(y)+T_0(x)T_2(y)}{\rt{(x^2-1)(y^2-1)}}
+\del_x^3\del_y\frac{T_3(x)T_1(y)+T_2(x)T_0(y)}{\rt{(x^2-1)(y^2-1)}}\right)\\
&-\frac{1}{48}\left(\del_x\del_y^2\frac{T_1(x)T_2(y)}{\rt{(x^2-1)(y^2-1)}}
+\del_x^2\del_y\frac{T_2(x)T_1(y)}{\rt{(x^2-1)(y^2-1)}}\right)\\
&+\frac{1}{192}\del_x^2\del_y^2\frac{T_2(x)T_2(y)+T_1(x)T_1(y)}{\rt{(x^2-1)(y^2-1)}}.
 \end{aligned}
\end{align}
For $(g,h)=(0,4)$ we find
\begin{align}
 \begin{aligned}
  W_{0,4}(x_1,\cdots,x_4)&=\frac{1}{32}
\Biggl[6+4\sum_{i<j}x_ix_j +3\Bigl(1+\sum_{i<j}x_ix_j+\prod_j x_j\Bigr)\sum_{k=1}^4(x_k^2-1)^{-1}\Biggr]
\prod_{j=1}^4(x_j^2-1)^{-3/2}
 \end{aligned}
\end{align}
which is rewritten as
\begin{align}
 \begin{aligned}
  &W_{0,4}(x_1,\cdots,x_4)\\
=&\frac{1}{16}\prod_{k=1}^4\del_k \Bigl(2\sum_{i<j}x_ix_j
+3\prod_j x_j\Bigr)\prod_{k=1}^4 (x^2_k-1)^{-1/2}\\
-&\frac{1}{32}\sum_{k=1}^4\del_k^2\prod_{l\ne k}\del_l 
\Biggl[T_2(x_k)\Bigl(1+\sum_{l_1<l_2}x_{l_1}x_{l_2}\Bigr)
+x_k\Bigl(\prod_{l}x_l+\sum_{l}x_{l}\Bigr)\Biggr]
\prod_{k=1}^4 (x^2_k-1)^{-1/2}.
 \end{aligned}
\label{eq:w04}
\end{align}
Finally, we find $\mathcal{J}_{1,2}(z)$ and $\mathcal{J}_{0,4}(z)$
in the form of the density integral \eqref{eq:Jgh-density}
\begin{align}
\begin{aligned}
\mathcal{J}_{1,2}(z) &=\frac{1}{\pi^2}\int \prod_{i=1}^2\frac{du_i}{\rt{1-u_i^2}}\Biggl[\frac{1}{96}
(T_1(u_1)T_3(u_2)+T_0(u_1)T_2(u_2))\del_1f(u_1,z)\del_2^3 f(u_2,z)\\
&\qquad\quad+\frac{1}{24}T_1(u_1)T_2(u_2)\del_1 f(u_1,z)\del_2^2f(u_2,z)\\
&\qquad\quad+\frac{1}{192}(T_2(u_1)T_2(u_2)+T_1(u_1)T_1(u_2))\del_1^2f(u_1,z)\del_2^2 f(u_2,z)\Biggr],
\end{aligned} 
\end{align}
and
\begin{align}
 \begin{aligned}
\mathcal{J}_{0,4}(z)&=\frac{1}{\pi^4}\int \prod_{i=1}^4\frac{du_i}{\rt{1-u_i^2}}
\Biggl[\frac{1}{16}\Bigl(2\sum_{i<j}u_iu_j+3\prod_i u_i\Bigr)\prod_{k=1}^4\del_k f(u_k,z)\\
&+\frac{1}{8} \Bigl[T_2(u_4)\Bigl(\sum_{l}u_{l}+\prod_l u_l\Bigr)+T_1(u_4)\Bigl(1+\sum_{l_1<l_2}u_{l_1}u_{l_2}\Bigr)\Bigr]\del_4^2f(u_4,z)\prod_{l=1}^3\del_l f(u_l,z)\Biggr] , 
 \end{aligned}
\label{eq:J04}
\end{align}
where $\del_i=\frac{\del}{\del u_i}$. The indices $i,j$ in the first line
of \eqref{eq:J04} run over $1,2,3,4$ and
$l,l_1,l_2$ in the second line run over $1,2,3$.
Again, one can check that this reproduces the small $\la$
expansion of
$J_3(z)$ in \eqref{eq:smallla-J}.

In a similar manner, one can in principle 
compute the $1/N$ corrections $J_n(z)$ in \eqref{eq:Jn-Jgh}
up to any desired order.

\section{Large $\la$ behavior and  a novel scaling limit}\label{sec:large-la}
In this section, we consider the large $\la$
behavior of the generating function $J(z)$
and the Wilson loop $W_{A_k}$ in the limit \eqref{eq:xlim}.
In principle, the large $\la$ behavior of $W_{A_k}$ is
obtained from that of $J(z)$ since they are related by the integral 
transformation
\begin{align}
 W_{A_k}=e^{-\frac{\la k^2}{8N^2}}\oint\frac{dz}{2\pi \ri z^{k+1}}e^{NJ(z)}
=e^{-\frac{\la x^2}{8}}\oint\frac{dz}{2\pi \ri z}e^{N[J(z)-x\log z]},
\label{eq:W-oint}
\end{align}
where $x=k/N$ defined in \eqref{eq:xlim},  
and the $1/N$ expansion of $W_{A_k}$ is also obtained from the expansion
of $J(z)$ in \eqref{eq:Jz-Jn}
\begin{align}
 \log W_{A_k}=\sum_{n=0}^\infty N^{1-n}S_n.
\label{eq:WAk-small}
\end{align}
The leading term $S_0$ is simply given by the Legendre transformation of $J_0(z)$
\begin{align}
 S_0=J_0(z_*)-x\log z_*,
\label{eq:S0-Legendre}
\end{align}
where $z_*$ is a solution of the saddle point equation
\begin{align}
 z\del_z J_0(z)\Bigl|_{z=z_*}=x.
\label{eq:saddle-z}
\end{align}
In the large $\la$ limit, 
the derivative of $J_0(z)$ in \eqref{eq:J0-wigner}
has a useful interpretation as a
system of fermions with temperature $1/\rt{\la}$
\begin{align}
 z\del_z J_0(z)=\frac{2}{\pi}\int_{-1}^1 du\rt{1-u^2}z\del_z f(u,z)
=\frac{2}{\pi}\int_{-1}^1 du\rt{1-u^2}\frac{1}{1+e^{\rt{\la}(\cos\th-u)}},
\label{eq:dJ0-int}
\end{align}
where we have set $z=e^{-\rt{\la}\cos\th}$ as in \eqref{eq:z-cos}.
Namely, the last factor $z\del_z f(u,z)$ in the integrand 
of \eqref{eq:dJ0-int} becomes the Fermi distribution function after setting
$z=e^{-\rt{\la}\cos\th}$.
In \cite{Horikoshi:2016hds}, the large $\la$ 
expansion of $S_0$ in \eqref{eq:S0-Legendre} was found by using the low temperature expansion
of Fermi distribution function,
known as the Sommerfeld expansion \footnote{There are typos in the
expansion of $S_0$ in \cite{Horikoshi:2016hds}, which were corrected in \cite{Gordon:2017dvy}.}
\begin{align}
\begin{aligned}
   S_0&=\frac{4\pi }{\la}\Biggl[\frac{(\rt{\la}\sin\th_k)^3}{6\pi^2}+\frac{\rt{\la}\sin\th_k}{12}\\
&\qquad-\frac{\pi^2(19+5\cos2\th_k)}{1440\rt{\la}\sin^3\th_k}
-\frac{\pi ^4 (6788 \cos 2 \th_k+35 \cos 4 \th_k+8985)}
{725760 \la^{3/2}\sin^7\th_k}+\cdots\Biggr],
 \end{aligned}
\end{align}
where $\th_k$ is given by 
\begin{align}
 \pi x=\th_k-\sin\th_k\cos\th_k.
\label{eq:def-thk}
\end{align}

To study the large $\la$ expansion of
higher order corrections $J_n(z)$,  
we can use the large $\la$ behavior of $f(u,z)$ in \eqref{eq:def-fu}
\begin{align}
 \begin{aligned}
  f(u,z)&=\rt{\la}(u-\cos\th)\Theta(u-\cos\th)+\mathcal{O}(\la^{-1/2}),\\
\del_u f(u,z)&=\rt{\la}\Theta(u-\cos\th)+\mathcal{O}(\la^{-1/2}),\\
\del_u^2 f(u,z)&=\rt{\la}\cob(u-\cos\th)+\mathcal{O}(\la^{-1/2}),
 \end{aligned}
\label{eq:f-approx}
\end{align}
where $z$ and $\th$ are related by \eqref{eq:z-cos}, and $\Theta(u)$ denotes the step function
\begin{align}
 \Theta(u)=\Biggl\{
\begin{aligned}
 1,&\qquad (u>0),\\
0,&\qquad(u<0).
\end{aligned}
\end{align} 
One can see from \eqref{eq:f-approx} that $\del_u^n f(u,z)$ behaves as $\la^{1/2}$ in the large $\la$ limit.
From the general structure of $\mathcal{J}_{g,h}$ in \eqref{eq:Jgh-density}, 
it follows that the large $\la$ expansion of $J_n(z)$ takes the form
 \begin{align}
 J_n(z)= \la^{\frac{n+1}{2}}\sum_{g=0}^\infty G_{g,n+1}(\th)\la^{-g}.
\label{eq:JtoG}
\end{align}
The leading term $G_{0,2}(\th)$ of the large $\la$ expansion of $J_1(z)$ was obtained in
\cite{Gordon:2017dvy}.
It would be interesting to find the large $\la$ expansion of $J_n(z)$ 
systematically along the lines of
\cite{Horikoshi:2016hds}.
\subsection{Scaling limit of the generating function and Wilson loops}
As suggested in \cite{Gordon:2017dvy}, we can take the scaling limit
\eqref{eq:scaling} in the regime $\la\sim N^2$. In this limit, the large $N$ expansion of $\log P(z)$
can be re-organized into the form of closed string genus expansion \eqref{eq:closed-expansion}.
Plugging the large $\la$ expansion of $J_n(z)$ \eqref{eq:JtoG}
into \eqref{eq:log-exp} and rewriting it 
in terms of the variable $\xi=\rt{\la}/N$, 
we find
\begin{align}
 \log P(z)=\sum_{n=0}^\infty N^{1-n}J_n(z)=\sum_{g=0}^\infty N^{2-2g}G_g(\xi,\th),
\end{align}
where $G_g(\xi,\th)$ is given by
\begin{align}
 G_g(\xi,\th)=\sum_{n=1}^\infty \xi^{n-2g}G_{g,n}(\th).
\end{align}

Now let us focus on the genus-zero part $G_0(\xi,\th)$
\begin{align}
 G_0(\xi,\th)=\sum_{n=1}^\infty \xi^n G_{0,n}(\th).
\end{align}
One can see that only $\mathcal{J}_{0,h}$ contributes to $G_{0,h}$
in this limit \eqref{eq:scaling}
\begin{align}
 \lim_{N,\la\to\infty, \xi=\frac{\rt{\la}}{N}}
\frac{N^{-h}\mathcal{J}_{0,h}}{h!}
=\xi^h G_{0,h}.
\end{align}
Then one can easily evaluate $G_{0,h}$
by plugging the approximation of $f(u,z)$ \eqref{eq:f-approx}
into the integral representation of $\mathcal{J}_{0,h}$ in \eqref{eq:Jgh-density}.
For instance, $G_{0,1}$ is given by
\begin{align}
 \begin{aligned}
  G_{0,1}=\frac{2}{\pi}\int_{\cos\th}^1\rt{1-u^2}(u-\cos\th)
=\frac{2}{\pi}
\left[\frac{\sin^3\th}{3}-\hf(\th-\sin\th\cos\th)\cos\th\right],
 \end{aligned}
\end{align}
and $G_{0,2}$ is given by
\begin{align}
\begin{aligned}
 G_{0,2}&=\frac{1}{8\pi^2} \int_{-1}^1 du\int_{-1}^1 dv\frac{(1-uv)\Bigl[(u-\cos\th)\Theta(u-\cos\th)-(v-\cos\th)\Theta(v-\cos\th)\Bigr]^2}{\rt{(1-u^2)(1-v^2)}(u-v)^2}\\
&=\frac{1}{8\pi^2}(\sin^2\th-\th\sin2\th+\th^2).
\end{aligned}
\end{align}
This reproduces the $1/N$ correction found in \cite{Gordon:2017dvy}.
We can proceed to higher orders in a similar manner
\begin{align}
 \begin{aligned}
  G_{0,3}&=
\frac{1}{48\pi^3}\int_{\cos\th}^1 du\int_{\cos\th}^1 dv\int_{\cos\th}^1 dw 
\frac{u+v+w+uvw}{\rt{(1-u^2)(1-v^2)(1-w^2)}}=\frac{1}{48\pi^3}(3\th^2\sin\th+\sin^3\th),
 \end{aligned}
\end{align}
and
\begin{align}
 G_{0,4}=\frac{1}{384\pi^2}
\Bigl[(2-\sin^2\th)\sin^2\th+3\th\sin2\th+6\th^2+2\th^3\cot\th\Bigr].
\end{align}

In this scaling limit \eqref{eq:scaling},
the Wilson loop $W_{A_k}$ has the closed string genus expansion as well 
\begin{align}
 \log W_{A_k}=\sum_{g=0}^\infty N^{2-2g}\mathcal{S}_g.
\label{eq:W-genus}
\end{align} 
The genus-zero term $\mathcal{S}_0$ is easily obtained from  
$G_0(\xi,\th)$ by
the Legendre transformation
\begin{align}
N^2\mathcal{S}_0 =N^2G_0(\xi,\th_*)-k\log z_*-\frac{\la k^2}{8N^2}=N^2
\Bigl[G_0(\xi,\th_*)+x\xi \cos\th_*-\frac{\xi^2 x^2}{8}\Bigr],
\end{align}
where $\th_*$ denotes the solution of the saddle point equation
\begin{align}
 \frac{\del}{\del\th}\bigl[G_0(\xi,\th)+x\xi \cos\th\bigr]\Bigl|_{\th=\th_*}=0.
\end{align}
From the result of $G_{0,h}~(h=1,\cdots,4)$ obtained above,
we find $\mathcal{S}_0$  as a power series in $\xi$
\begin{align}
 \mathcal{S}_0=\sum_{n=1}^\infty s_n\xi^n,
\end{align}
with
\begin{align}
  s_1=\frac{2\sin^3\th_k}{3\pi},\quad
s_2=\frac{\sin^4\th_k}{8\pi^2},\quad
s_3=\frac{\sin^3\th_k}{48\pi^3},\quad
s_4=\frac{(2-\sin^2\th_k)\sin^2\th_k}{384\pi^4},
\label{eq:result-sn}
\end{align}
where $\th_k$ is defined in \eqref{eq:def-thk}.
Interestingly, the polynomial dependence on $\th$
in $G_{0,h}$ cancels after performing the Legendre transformation, and the resulting
expression of $s_n$ is a trigonometric function of $\th_k$.
This was observed for $s_2$ in \cite{Gordon:2017dvy}
and we believe that this is true for all $s_n$.
It would be nice to find a general proof of this statement.\footnote{
Since the Wilson loop $W_{A_k}$ in $SU(N)$ theory
is invariant under $k\to N-k$, it follows that $\mathcal{S}_0$ is invariant 
under $\th_k\to\pi-\th_k$. However,
this invariance alone is not strong enough to prove this statement. 
We would like to thank the anonymous referee of JHEP for a comment on this point.}

\begin{figure}[htb]
\centering
\subcaptionbox{$s_1\xi$  and $s_1\xi+s_2\xi^2$\label{sfig:xi2}}{\includegraphics[width=4.6cm]{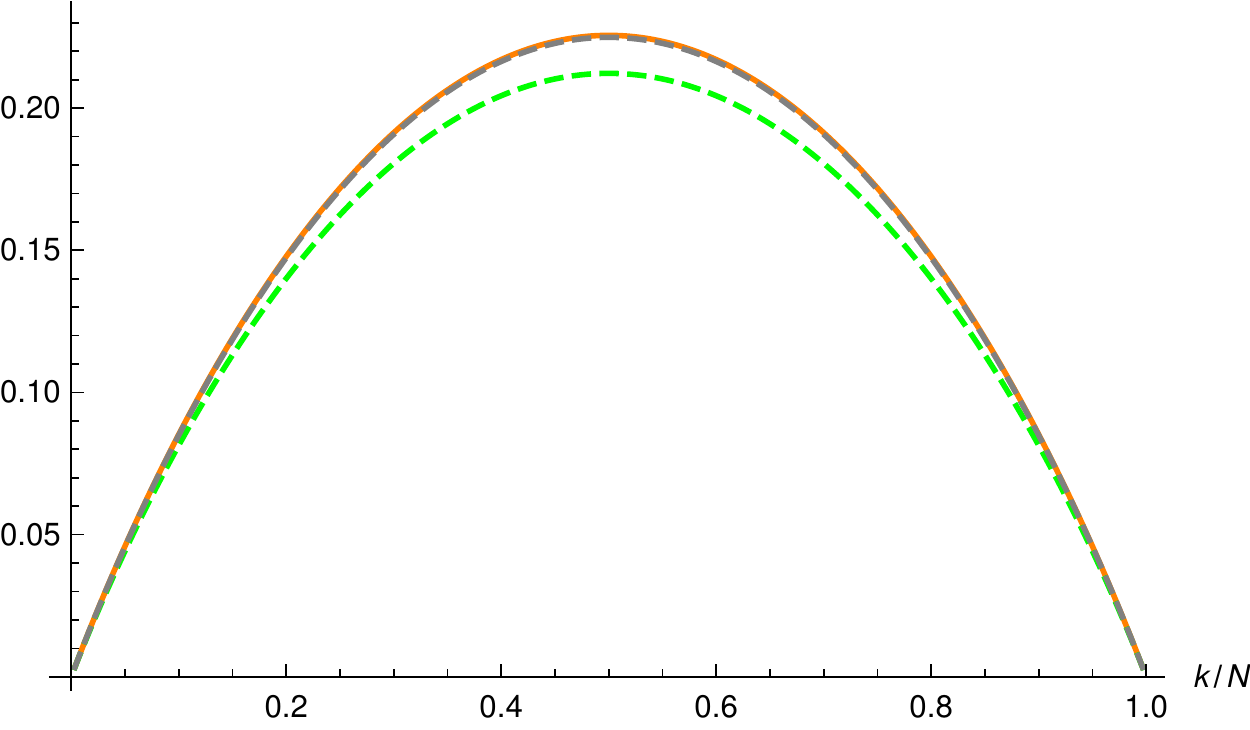}}
\hskip3mm
\subcaptionbox{$s_2\xi^2$ and $s_2\xi^2+s_3\xi^3$ \label{sfig:xi3}}{\includegraphics[width=4.6cm]{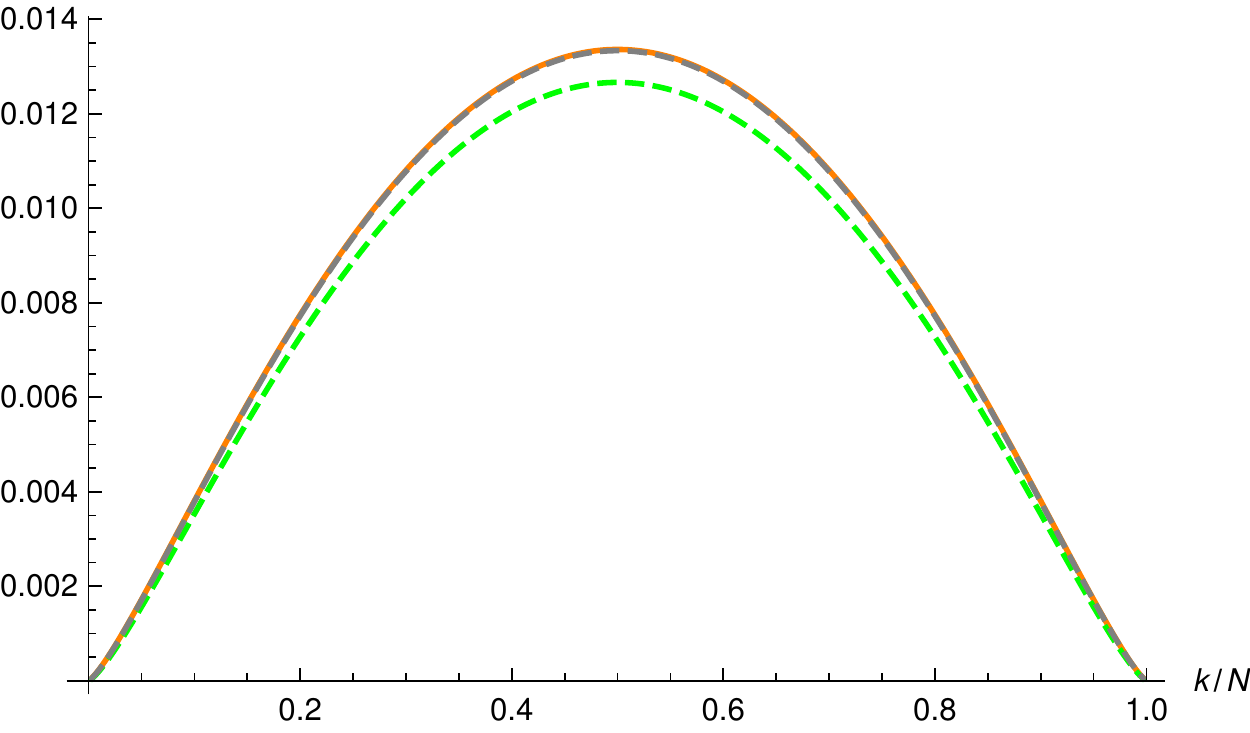}}
\hskip3mm
\subcaptionbox{$s_3\xi^3$ and $s_3\xi^3+s_4\xi^4$ \label{sfig:xi4}}{\includegraphics[width=4.6cm]{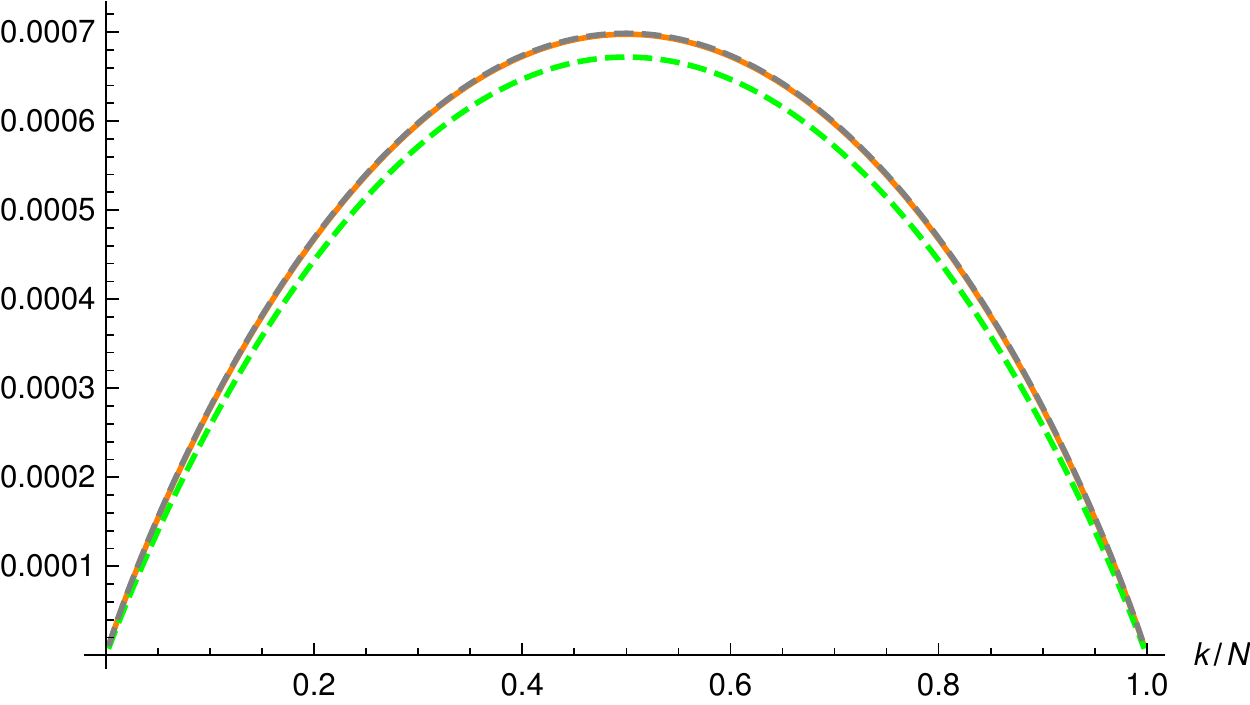}}
  \caption{
Plot 
of $\frac{1}{N^2}\log W_{A_k}$ as a function of $k/N$ for $N=300,\xi=1$.
In \subref{sfig:xi2}, the orange curve represents
the exact value of $\frac{1}{N^2}\log W_{A_k}$, while
the green  and the gray dashed curves represent
$s_1\xi$ and $s_1\xi+s_2\xi^2$, respectively.
In \subref{sfig:xi3} and \subref{sfig:xi4}, we show the plot of
$\frac{1}{N^2}\log W_{A_k}-s_1\xi$ and $\frac{1}{N^2}\log W_{A_k}-s_1\xi-s_2\xi^2$, respectively. 
Again, the orange curve is the exact value, while the green and the gray dashed curves represent
the next order term and the sum of the next and the next-to-next order terms, respectively. 
}
  \label{fig:W-scale}
\end{figure}

We can compare this result \eqref{eq:result-sn}
with the exact values of $W_{A_k}$ in \eqref{eq:W-exact}
for some fixed $\xi=\rt{\la}/N$ with a large $N$.
As we can see from Fig.~\ref{fig:W-scale},
our result \eqref{eq:result-sn} agrees nicely with the exact value of $W_{A_k}$.
For example, in Fig.~\ref{sfig:xi3} we show the plot of 
the exact value of $\frac{1}{N^2}\log W_{A_k}-s_1\xi$ (orange curve), $s_2\xi^2$ (green dashed curve),
and $s_2\xi^2+s_3\xi^3$ (gray dashed curve) for $N=300$ and $\xi=1$.
One can clearly see that the orange curve and the gray dashed curve match quite well,
which confirms our result of $s_3$ in \eqref{eq:result-sn}.
\subsection{Resolvent in the scaling limit}
In the scaling limit \eqref{eq:scaling},
the matrix model potential \eqref{eq:V-mod} becomes 
\begin{align}
 V(w)=2w^2-\xi(w-\cos\th)\Theta(w-\cos\th),
\label{eq:potential}
\end{align}
where $w$ denotes the eigenvalue of the matrix $M$.
\begin{figure}[htb]
\centering
\subcaptionbox{$\xi=2$\label{sfig:p-xi2}}{\includegraphics[width=4.6cm]{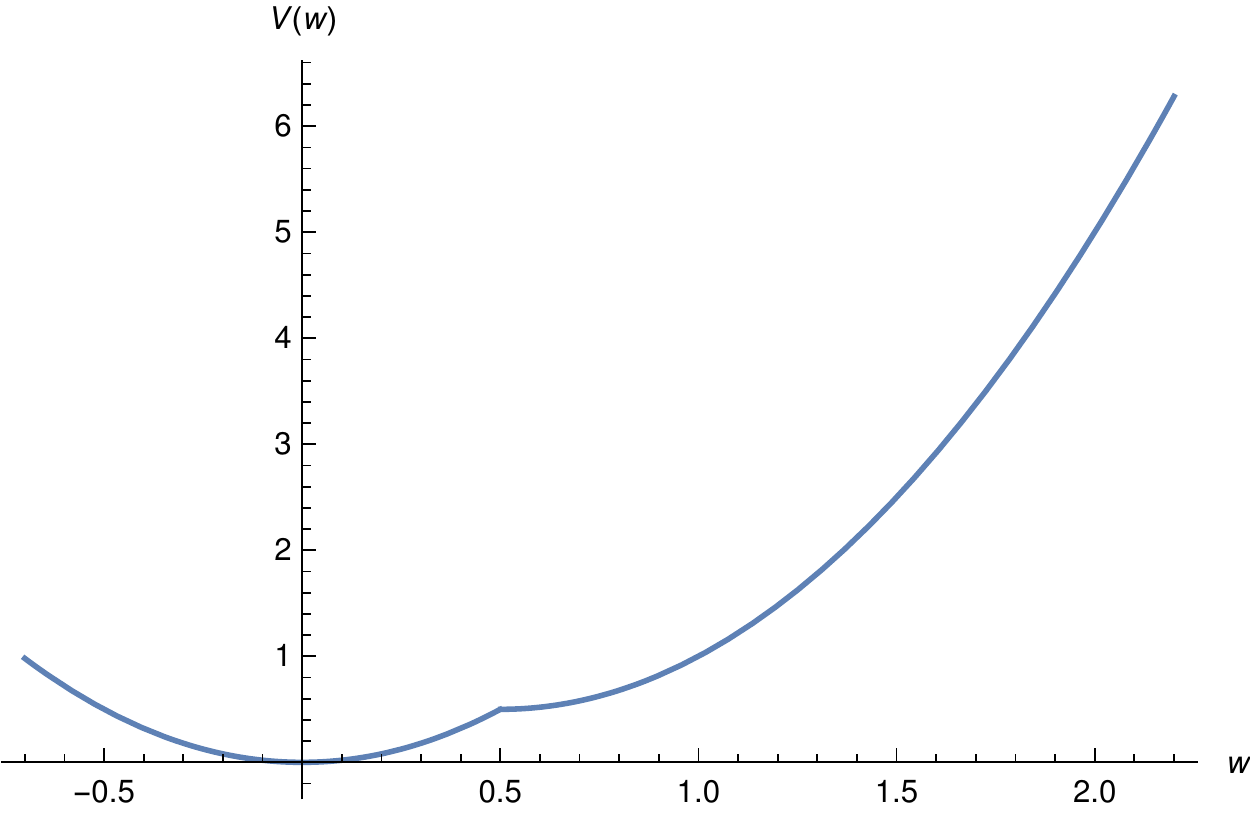}}
\hskip5mm
\subcaptionbox{$\xi=3$ \label{sfig:p-xi3}}{\includegraphics[width=4.6cm]{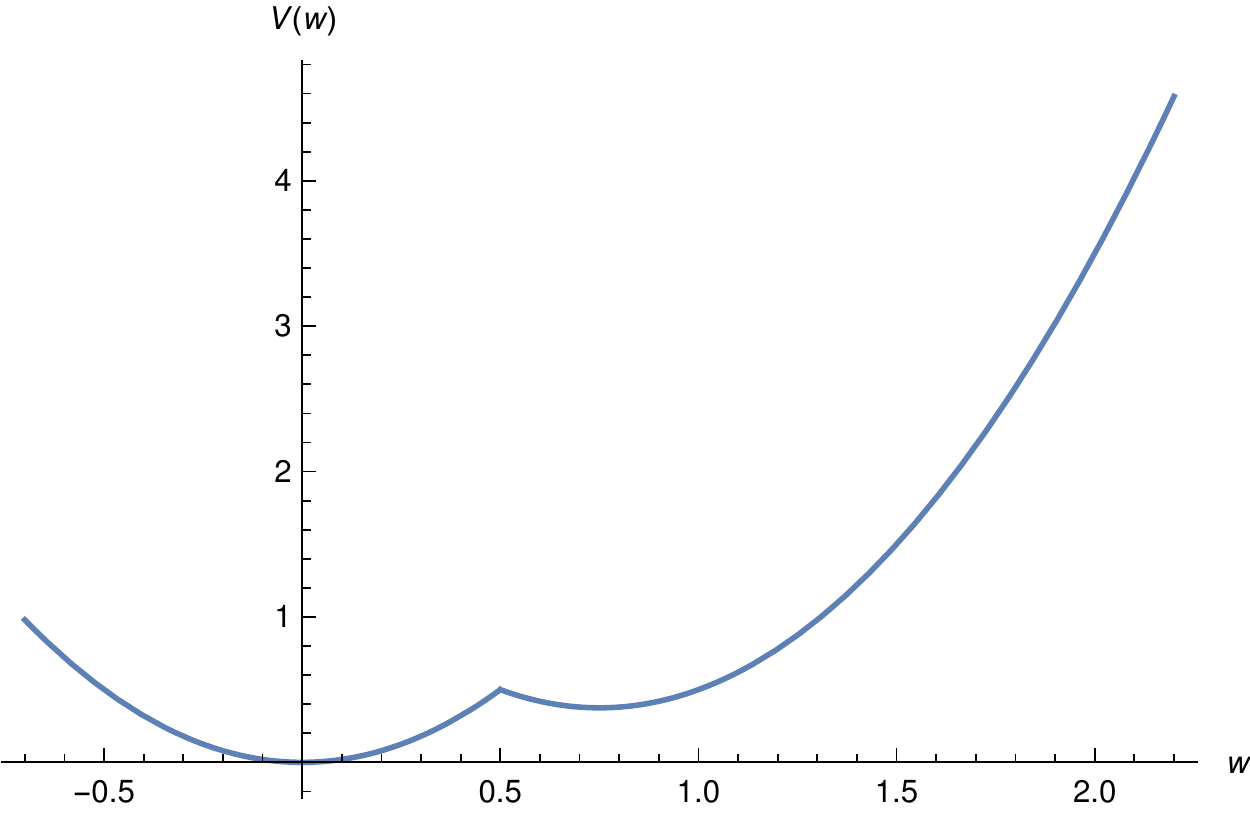}}
\hskip5mm
\subcaptionbox{$\xi=5$ \label{sfig:p-xi5}}{\includegraphics[width=4.6cm]{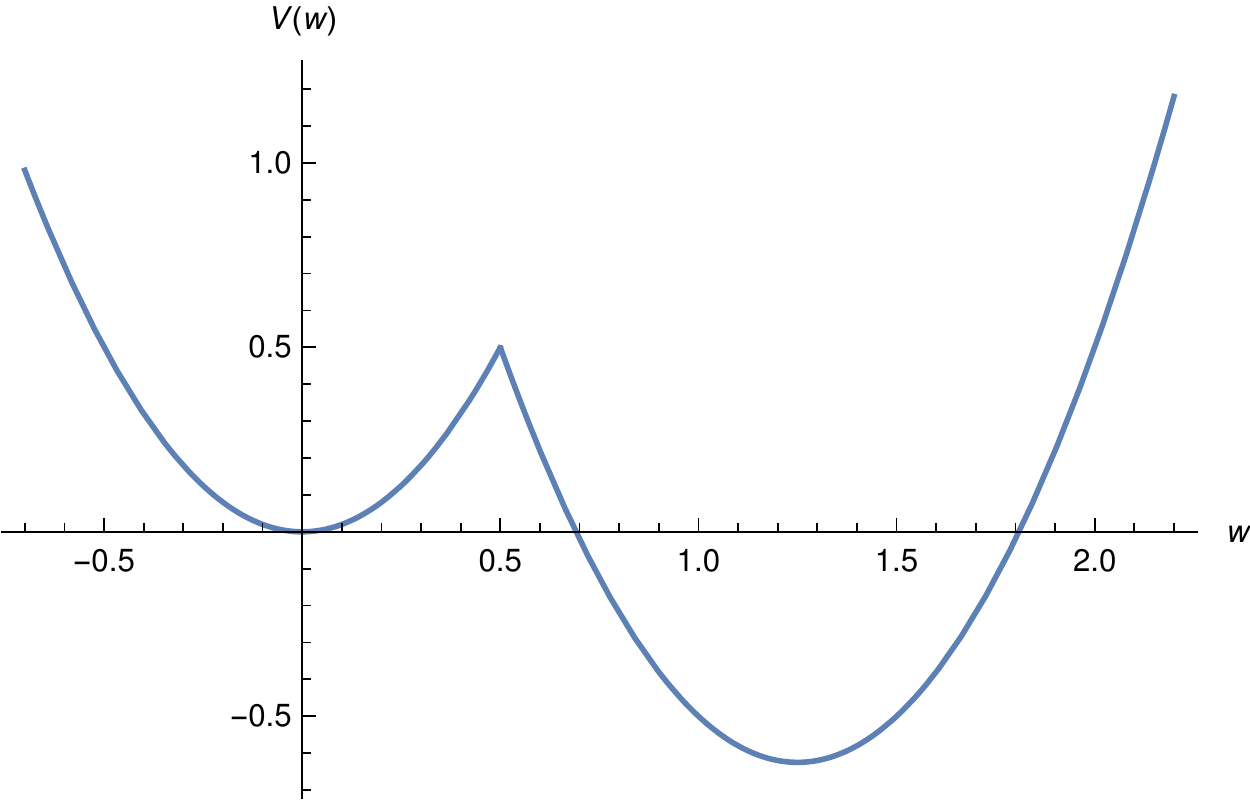}}
  \caption{
Plot 
of the potential $V(w)$ in \eqref{eq:potential} for $\th=\pi/3$ with
\subref{sfig:p-xi2} $\xi=2$, \subref{sfig:p-xi3} $\xi=3$, and
\subref{sfig:p-xi5} $\xi=5$.
}
  \label{fig:pot}
\end{figure}
As we increase the value of $\xi$, this potential
changes as in Fig.~\ref{fig:pot}.
In particular, at some critical value $\xi=\xi_c$
the potential develops a new local minimum other than the original minimum
of Gaussian potential near $w=0$.
From this figure, it is natural to expect that when $\xi<\xi_c$
the eigenvalues are supported near the original minimum $w=0$, while
when $\xi>\xi_c$ the eigenvalues are also distributed around the new minimum
and the eigenvalue density has two supports. In other words,
there is a phase transition from the one-cut phase to
the two-cut phase at some critical point $\xi_c$.
In this subsection, we construct the planar solution of resolvent in the
one-cut phase and study its behavior as we increase the value of $\xi$.

The planar resolvent in the one-cut phase is given by
the general formula \cite{Migdal:1984gj}
\begin{align}
\begin{aligned}
 R(z)&=\int_a^b du\frac{\rho(u)}{z-u}=\int_C\frac{dw}{4\pi\ri}\frac{V'(w)}{z-w}
\rt{\frac{(z-a)(z-b)}{(w-a)(w-b)}},
\end{aligned}
\label{eq:res-general}
\end{align}
where $C$ is a contour enclosing the cut $[a,b]$ and
$V'(w)$ is given by
\begin{align}
 V'(w)=4w-\xi \Theta(w-\cos\th).
\end{align}
Although this is a non-analytic function,
we can define the step function as a limit of the Fermi distribution
function
\begin{align}
 \Theta(w)=\lim_{\varepsilon\to+0}\frac{1}{1+e^{-\frac{w}{\varepsilon}}},
\end{align}
and we use the general formula \eqref{eq:res-general}
in this sense.
Assuming that the point $w=\cos\th$ is
located inside the cut $[a,b]$ 
\begin{align}
 a<\cos\th<b,
\end{align} 
this integral \eqref{eq:res-general} can be evaluated as
\footnote{A similar computation
has appeared in \cite{Morita:2017oev}
in the context of Chern-Simons matrix models.}
\begin{align}
\begin{aligned}
 R(z)&=\int_C\frac{dw}{4\pi\ri}\frac{4w}{z-w}
\rt{\frac{(z-a)(z-b)}{(w-a)(w-b)}}-
\int_{\cos\th}^b \frac{dw}{2\pi\ri}\frac{\xi}{z-w}
\rt{\frac{(z-a)(z-b)}{(w-a)(w-b)}}\\
&=2z-2\rt{(z-a)(z-b)}-\frac{\xi}{\pi}\text{arctan}\left(m\rt{\frac{z-a}{z-b}}\right),
\end{aligned}
\label{eq:res-ab}
\end{align}
where $m$ is defined by
\begin{align}
 m=\rt{\frac{b-\cos\th}{\cos\th-a}}.
\label{eq:m-def}
\end{align}
Requiring the following large $z$ behavior
of the resolvent
\begin{align}
 \lim_{z\to\infty}R(z)=r_0+\frac{r_1}{z}+\mathcal{O}(z^{-2}),\quad r_0=0,\quad r_1=1,
\end{align}
we find the condition for the end-points of cut
\begin{align}
 \begin{aligned}
  a+b-\frac{\xi}{\pi}\text{arctan}(m)&=0,\\
\qu(a-b)^2+\frac{\xi}{2\pi}\frac{m}{m^2+1}(a-b)&=1.
 \end{aligned}
\end{align}
These conditions are solved as
\begin{align}
 a=f_1(m)-f_2(m),\quad
b=f_1(m)+f_2(m),
\label{eq:ab-fm}
\end{align}
where we introduced the functions $f_1(m)$ and $f_2(m)$ by
\begin{align}
\begin{aligned}
 f_1(m)&=\frac{\xi}{2\pi}\text{arctan}(m) ,\\
f_2(m)&=\al+\rt{1+\al^2},\qquad
\al=\frac{\xi}{2\pi}\frac{m}{m^2+1}.
\end{aligned}
\end{align}
Finally, plugging the solution of $a,b$ \eqref{eq:ab-fm} into \eqref{eq:m-def},
we find the equation for $m$
\begin{align}
 (1+m^2)(f_1(m)-\cos\th)+(1-m^2)f_2(m)=0.
\end{align}
This fixes $m$ as a function of $\xi$ and $\th$,
and via the relation \eqref{eq:ab-fm} $a$ and $b$ also become
functions of $\xi$ and $\th$.

Let us first consider the behavior of this one-cut solution in the 
small $\xi$ limit.
When $\xi=0$ one can easily see that $a,b$ and $m$ are given by
\begin{align}
 a=-1,\quad b=1,\quad m=\tan\frac{\th}{2}.
\label{eq:abm-zero}
\end{align}
Then we can easily find the small $\xi$ expansion of $a,b$ and $m$
around these values in \eqref{eq:abm-zero}
\begin{align}
\begin{aligned}
 a&=-1+\frac{\xi}{4\pi}(\th-\sin\th)+\mathcal{O}(\xi^2),\\
b&=1+\frac{\xi}{4\pi}(\th+\sin\th)+\mathcal{O}(\xi^2),\\
m&=\tan\frac{\th}{2}+\frac{\xi}{2\pi}\frac{(\th+\sin\th\cos\th)\sin^2\frac{\th}{2}}{\sin^3\th}
+\mathcal{O}(\xi^2). 
\end{aligned}
\label{eq:abm-small}
\end{align} 
Using \eqref{eq:abm-small}, the resolvent \eqref{eq:res-ab} is expanded as
\begin{align}
 R(z)=2z-2\rt{z^2-1}+\frac{\xi}{2\pi}\left[\frac{\th z+\sin\th}{\rt{z^2-1}}
-2\arctan\left(\tan\frac{\th}{2}\rt{\frac{z+1}{z-1}}\right)\right]+\mathcal{O}(\xi^2).
\end{align}
We can check that the order $\mathcal{O}(\xi)$ term of $R(z)$
reproduces the $1/N$
correction of resolvent found in \cite{Gordon:2017dvy}.

\begin{figure}[htb]
\centering
\subcaptionbox{$\xi=0.5$\label{sfig:rho-xi1o2}}{\includegraphics[width=4.6cm]{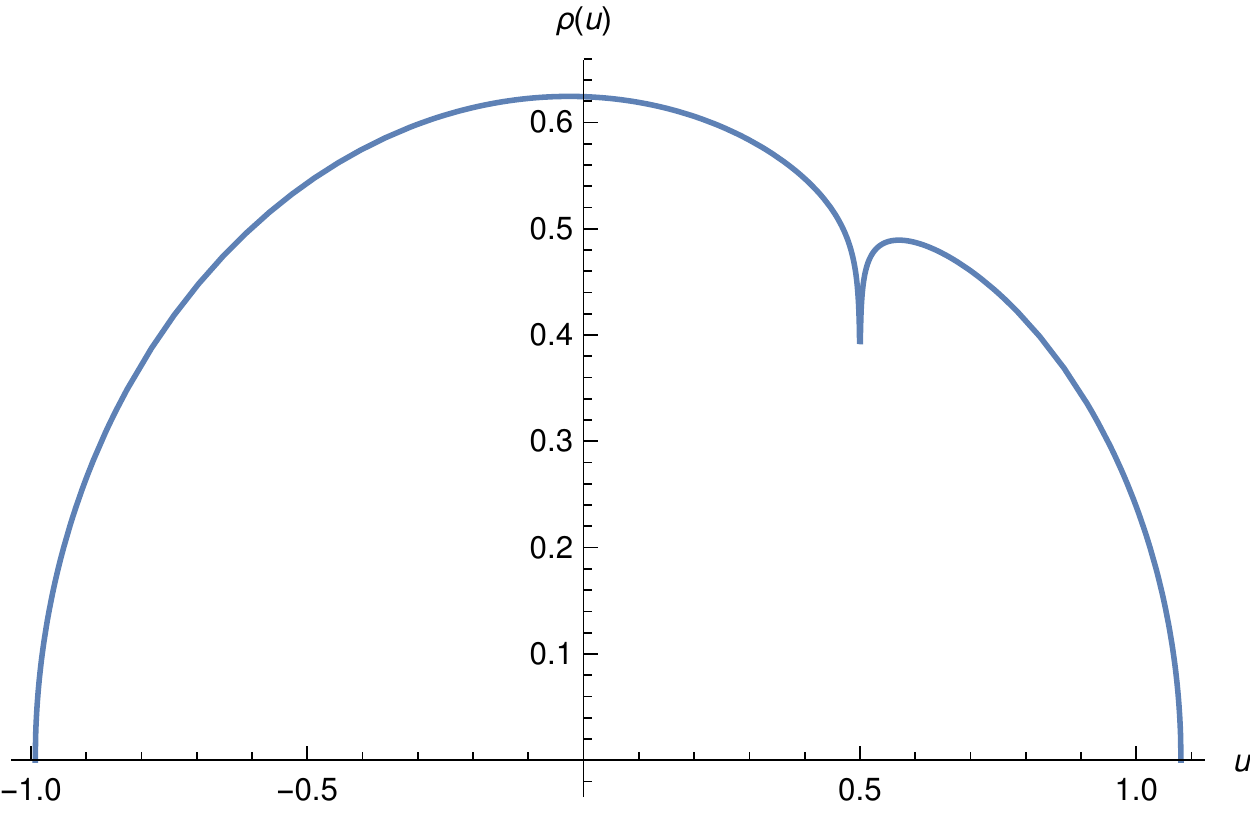}}
\hskip3mm
\subcaptionbox{$\xi=1$ \label{sfig:rho-xi1}}{\includegraphics[width=4.6cm]{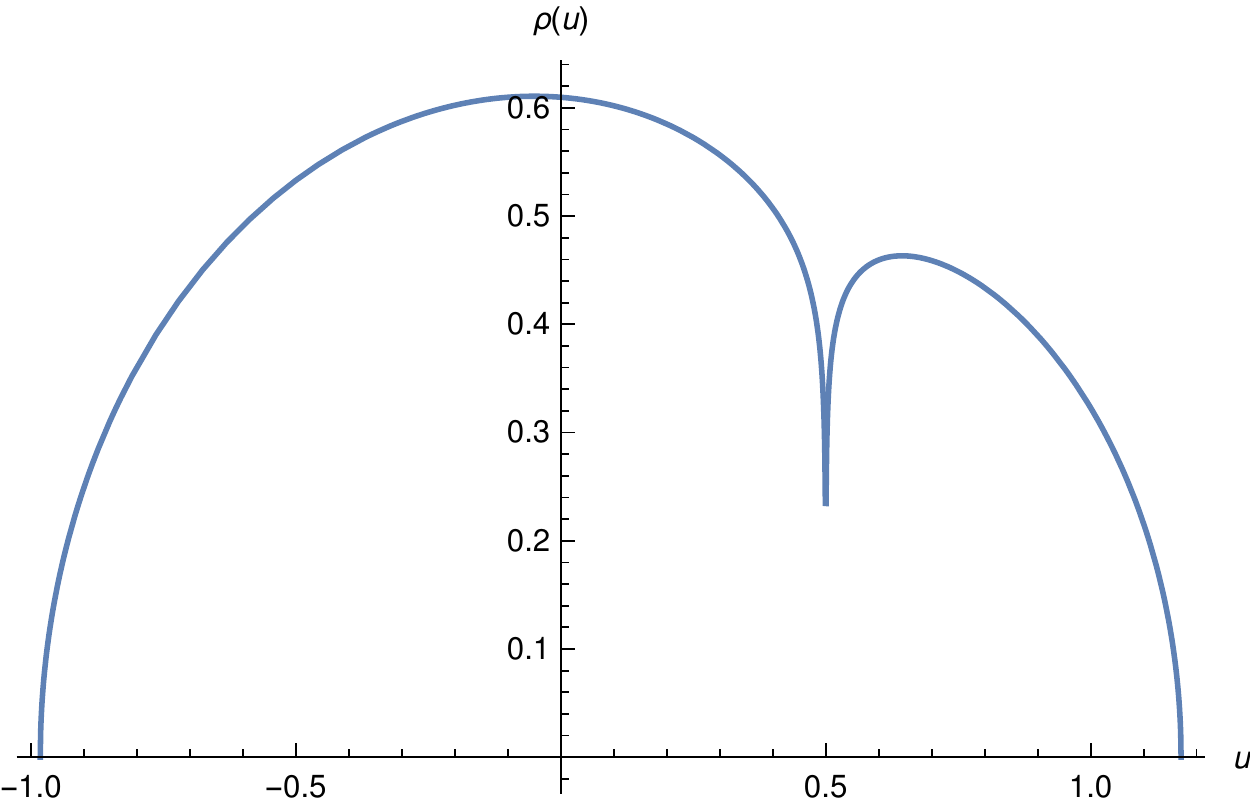}}
\hskip3mm
\subcaptionbox{$\xi=2$ \label{sfig:rho-xi2}}{\includegraphics[width=4.6cm]{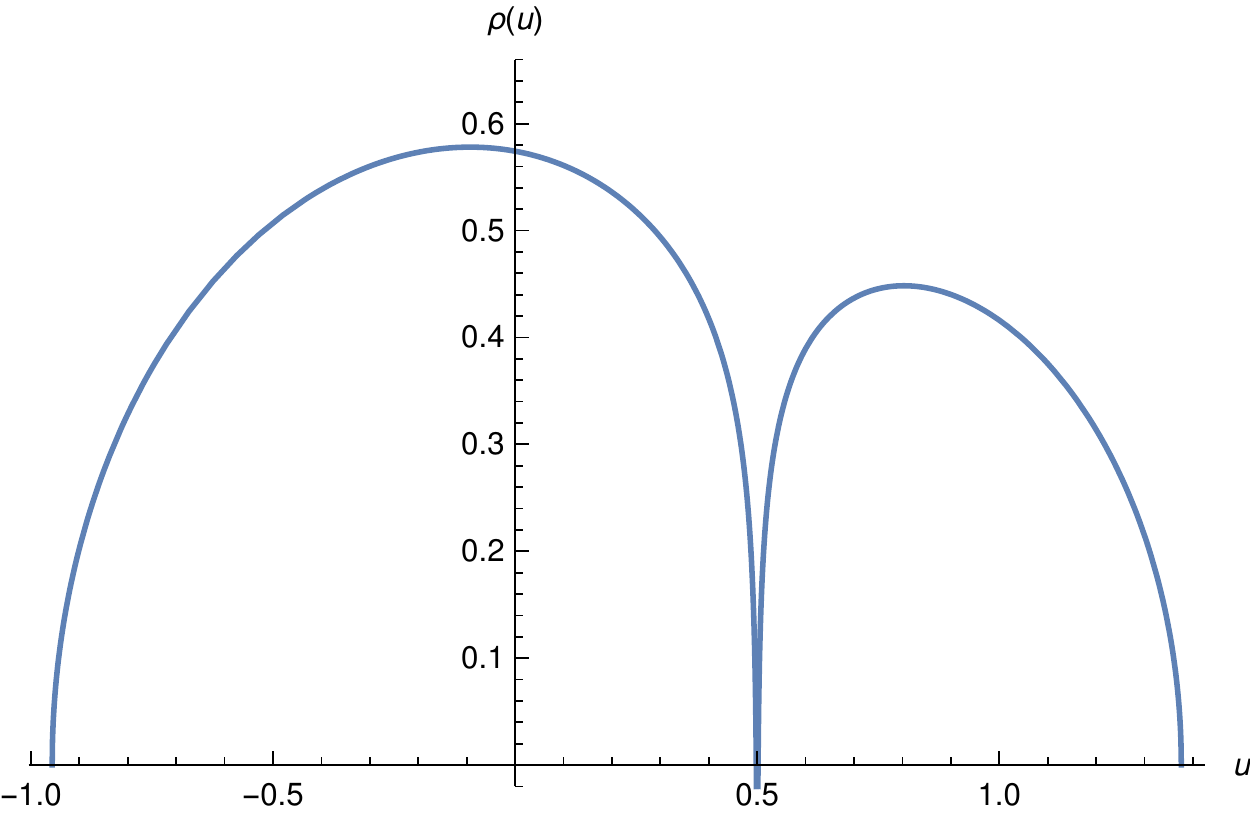}}
  \caption{
Plot 
of the eigenvalue density $\rho(u)$ in \eqref{eq:potential} for $\th=\pi/3$ with
\subref{sfig:rho-xi1o2} $\xi=0.5$, \subref{sfig:rho-xi1} $\xi=1$, and
\subref{sfig:rho-xi2} $\xi=2$.
}
  \label{fig:rho}
\end{figure}

The eigenvalue density $\rho(u)$ at {\it finite} $\xi$ can
also be obtained from the resolvent \eqref{eq:res-ab} by taking the discontinuity
across the cut $[a,b]$
\begin{align}
 \rho(u)=\frac{2}{\pi}\rt{(u-a)(b-u)}-\frac{\xi}{\pi^2}\text{arctanh}\left(m\rt{\frac{u-a}{b-u}}\right).
\label{eq:rho-abm}
\end{align}
Although we do not have a closed form
expression of $a,b$ and $m$ at finite value of $\xi$,
it is easy to compute them numerically. 
In Fig.~\ref{fig:rho}, we show the plot of the eigenvalue density
$\rho(u)$ for several values of $\xi$, with $\th=\pi/3$ as an example.
As we can see from Fig.~\ref{fig:rho},
when $\xi$ is small the density  $\rho(u)$
can be regarded as a small perturbation of the semi-circle
distribution of Gaussian matrix model.
As we increase the value of $\xi$, 
the eigenvalue density near $u=\cos\th$ decreases.
One can imagine that eventually the support of eigenvalue density splits into two parts
above some critical value $\xi>\xi_c$.

One can also compute the planer free energy 
using the eigenvalue density $\rho(u)$ in \eqref{eq:rho-abm}
\begin{align}
 G_0(\xi,\th)=-\int_a^b du \rho(u)V(u)+\hf \int_a^b du \int_a^b dv \rho(u)\rho(v)\log(u-v)^2 
-\frac{1}{N^2}\log Z,
\label{eq:G0-int}
\end{align}
where $Z$ is the partition function of Gaussian matrix model.
We have checked numerically that when $\xi$ is small
\eqref{eq:G0-int} agrees with the result of previous subsection
$G_0(\xi,\th)\approx \sum_{h=1}^4 \xi^h G_{0,h}(\th)$.

\section{Discussion}\label{sec:discussion}
In this paper we have argued that the one-cut phase
of the matrix integral \eqref{eq:V-mod} is smoothly
connected to the small $\la$ regime and we have 
demonstrated that the result of \cite{Gordon:2017dvy}
is correctly reproduced from the perturbative
computation in the Gaussian matrix model.
The higher order corrections in the $1/N$ expansion can be obtained
systematically by using the topological recursion in the Gaussian
matrix model.

In the scaling limit \eqref{eq:scaling},
the $1/N$ expansion of $W_{A_k}$ (or the generating function
thereof) takes the form
of the genus expansion of closed string.
This suggests that the Wilson loop $W_{A_k}$ in this limit
corresponds to a closed string background without D-branes on the dual bulk side. 
This is similar in spirit to 
the bubbling geometry dual to Wilson loops in large representations
studied in \cite{Yamaguchi:2006te,Lunin:2006xr,DHoker:2007mci,Okuda:2008px,Aguilera-Damia:2017znn},
where the Wilson loops are replaced by a pure geometric background whose topology is different
from the original $AdS_5\times S^5$.
However, there is a crucial difference between our case and  
the bubbling geometry considered 
in \cite{Yamaguchi:2006te,Lunin:2006xr,DHoker:2007mci,Okuda:2008px,Aguilera-Damia:2017znn}.
In the case of bubbling geometry we take a limit where 
the number of boxes in the Young diagram labeling the representation of Wilson
loop scales as $\mathcal{O}(N^2)$, while in our case of $W_{A_k}$ the number of boxes in the Young diagram is $k\sim \mathcal{O}(N)$.
At present it is not clear to us whether the closed string background
corresponding to the anti-symmetric Wilson loop $W_{A_k}$ in this scaling limit has a 
classical bulk gravity interpretation or not.

Here we speculate possible interpretation of the closed string background
appearing in the limit \eqref{eq:scaling}.
One possibility is that our case might correspond to 
a highly degenerate limit of the bubbling geometry where the size of ``bubbles'' 
(or non-trivial cycles in bubbling geometry)
become less than the string scale and it cannot be described by a classical solution of supergravity.
Another possibility is that in the limit 
\eqref{eq:scaling} we should go to the S-dual picture and the D5-branes
are replaced by the NS5-branes. 
This comes from the observation that our limit
\eqref{eq:scaling} can be thought of as the 't Hooft limit of S-dual theory\footnote{This was pointed out by Martin Kruczenski
during the workshop ``Localization and Holography'' at the University of Michigan, October 2017.
We would like to thank him for sharing his insight.}
\begin{align}
 \xi^{-2}=\frac{N^2}{\la}=\frac{N^2}{g_{\text{YM}}^2N}=\til{g}_{\text{YM}}^2N,\qquad
\til{g}_{\text{YM}}^2=\frac{1}{g_{\text{YM}}^2},
\end{align}
and in this S-dual picture the NS5-branes might be treated as a closed string background. 
We should emphasize that there might be other possibilities
and  we do not have a clear picture of the bulk side yet.  

In any case, it is important to understand 
the bulk interpretation better. We hope that the better understanding of the bulk side 
might also shed light on the
issue of discrepancy at the subleading order in $1/N$, mentioned in section \ref{sec:intro}.
 
In this paper we have also considered the behavior of $W_{A_k}$
as a function of $\xi$ in the scaling limit \eqref{eq:scaling}.
As we increase $\xi$ the potential $V(w)$
in \eqref{eq:potential} develops a new local minimum 
(see Fig.~\ref{fig:pot}),
and it is natural to
conjecture that there is a phase transition
between the one-cut phase and the two-cut phase at some critical value
$\xi=\xi_c$. Above the critical value $\xi>\xi_c$
the backreaction of operator insertion can no longer 
be ignored.
In this paper we have only considered the one-cut solution.
It would be very interesting to find the explicit
form of two-cut solution and study the nature of this phase transition, 
say, the order of phase transition. 
We leave this as an important future problem.

We should stress that the closed string expansion
\eqref{eq:closed-expansion} is expected to hold in both phases
in the scaling limit \eqref{eq:scaling}.
This suggests that the phase transition on the matrix model side has
a counterpart on the bulk string theory side.
It is well-known that, in the case of $\mathcal{N}=4$ SYM at finite temperature,
the Hagedorn transition on the field theory 
side corresponds to the Hawking-Page transition on the bulk
side, where two different geometries
exchange dominance in the bulk gravity path integral \cite{Witten:1998zw}.
However, in our case, it is not clear whether there is such an interpretation
of the exchange of dominance of two geometries (or two different closed string backgrounds).
It would be very interesting to understand the bulk interpretation of this phase transition
for $W_{A_k}$.

\vskip8mm
\centerline{\bf Acknowledgments}
\vskip2mm
\noindent
I would like to thank Matteo Beccaria for  collaboration at
the initial stage of this work.
I would also like to thank Jamie Gordon and Oleg Lunin
for correspondence.
This work  was supported in part by JSPS KAKENHI Grant Number 16K05316.

\appendix

\section{Small $\la$ expansion of $W_{A_k}$}\label{app:small-la}
In this appendix, we consider the small $\la$ expansion
of $W_{A_k}$.
This can, in principle, be computed by the usual Feynman diagram
expansion in the Gaussian matrix model.
However, a more efficient method is 
to extract the coefficient of $z^k$ from the exact result of the generating function $P(z)$
in \eqref{eq:P-exact}, 
and expand it around $\la=0$.
It turns out that the coefficient of $(\la/N)^n$
is a polynomial of $k$ and $N$,
and this polynomial can be found by 
using the {\tt InterpolatingPolynomial}
in {\tt Mathematica} from the data of
the small values of $k,N=1,2,\cdots$.

Using this method, we find the 
small $\la$ behavior of the $1/N$
expansion of $\log W_{A_k}$
in \eqref{eq:WAk-small}.
The first few terms are given by
\begin{align}
\begin{aligned}
 S_0&=d_0+\frac{a}{8}\la-\frac{a}{384}\la^2+\frac{a}{9216}\la^3-\frac{a (a+4)}{737280}\la^4+\frac{a (10 a+13)}{44236800}\la^5-\frac{a \left(40 a^2+774 a+495\right)}{29727129600}\la^6+\cdots,\\
 S_1&=d_1+\frac{a\la}{8}-\frac{a}{384}\la^2+\frac{a}{9216}\la^3-\frac{a (a+3)}{737280}\la^4+\frac{a (10 a+3)}{44236800}\la^5-\frac{a \left(40 a^2+548 a-279\right)}{29727129600}\la^6+\cdots,\\
S_2&=d_2+\frac{a}{737280}\la^4-\frac{7 a}{44236800}\la^5+\frac{a (226 a+229)}{29727129600}\la^6
+\cdots,\\
S_3&=d_3+\frac{a}{14745600}\la^5-\frac{19 a}{1415577600}\la^6+\cdots,
\end{aligned}
\label{eq:S0-exp}
\end{align}
where we defined $a$ by
\begin{align}
 a=x(1-x),\qquad x=k/N.
\end{align}
The $\la$-independent term $d_n$ in \eqref{eq:S0-exp} comes from the $1/N$ expansion
of the dimension $d_{A_k}=\frac{N!}{k!(N-k)!}$ of the anti-symmetric representation $A_{k}$
\begin{align}
 \log d_{A_k}=\sum_{n=0}^\infty N^{1-n}d_n,
\label{eq:dn}
\end{align}
which can be obtained as
\begin{align}
\begin{aligned}
 \log d_{A_k}&=\log\frac{\Ga(N+1)}{\Ga(Nx+1)\Ga(N-Nx+1)}\\
&=N\Bigl[-x\log x-(1-x)\log(1-x)\Bigr]-\hf\log\bigl[2\pi N x(1-x)\bigr]\\
&\qquad +\frac{1}{12N}\left[1-\frac{1}{x}-\frac{1}{1-x}\right]-\frac{1}{360N^3}\left[1-\frac{1}{x^3}
-\frac{1}{(1-x)^3}\right]+\cdots. 
\end{aligned}
\end{align}

In studying $W_{A_k}$,
one could normalize it by the dimension $d_{A_k}$,
but in this paper we have used the {\it un-normalized} Wilson loops.
Namely, in our definition of
$W_{A_k}$ we do not divide it by the dimension $d_{A_k}$.
This definition is important for our discussion of  
the scaling limit \eqref{eq:scaling},
since the $1/N$
expansion of $\log d_{A_k}$ is independent of $\la$
and it does not fit into the form 
of the closed string expansion \eqref{eq:W-genus}
in this limit \eqref{eq:scaling}.

We note in passing that 
the $\mathcal{O}(a)$ term of $S_0$ in 
\eqref{eq:S0-exp} is given by the planar result of 
Wilson loop in the fundamental representation \cite{Erickson:2000af}
\begin{align}
 S_0=d_0+a\log \left[\frac{2I_1(\rt{\la})}{\rt{\la}}\right]+\mathcal{O}(a^2).
\end{align}
It would be interesting find the closed form expression of the higher 
order terms in $a$.

From \eqref{eq:W-oint},
$W_{A_k}$ with fixed $k$ can also be computed once we know the generating function $J(z)$. 
At the leading order in $1/N$, the integral 
\eqref{eq:W-oint} is evaluated by the saddle point approximation, 
and the leading term $S_0$ is just
given by the Legendre transformation of
$J_0(z)$ \eqref{eq:S0-Legendre}.
From the small $\la$ expansion of $J_0(z)$ in \eqref{eq:smallla-J}
we can easily find the small $\la$ expansion of the
solution $z_*$
of the saddle point equation \eqref{eq:saddle-z}
\begin{align}
 z_*=\frac{x}{1-x}+\frac{2 x^2-x}{8 (1-x)}\la+\frac{6 x^3-7 x^2+2 x}{192 (1-x)}\la^2+\cdots.
\end{align}
Plugging this $z_*$ into \eqref{eq:S0-Legendre}, 
the leading term $S_0$ becomes
\begin{align}
\begin{aligned}
 S_0
&=-x\log x-(1-x)\log(1-x)+\frac{x(1-x)}{8}\la-\frac{x(1-x)}{384}\la^2+\cdots,
\end{aligned}
\end{align} 
which reproduces the small $\la$ expansion of $S_0$ 
in \eqref{eq:S0-exp},
as expected. 

In general, it is easier to study the $1/N$ expansion of the generating function $J(z)$
rather than directly analyzing the $1/N$ expansion of $\log W_{A_k}$.
These two approaches are related by the integral transformation \eqref{eq:W-oint}, of course.

 
\end{document}